\documentclass[graybox,vecarrow]{svmult}

\usepackage{mathptmx}       
\usepackage{helvet}         
\usepackage{courier}        
\usepackage{type1cm}        
\usepackage{makeidx}         
\usepackage{graphicx}        
\usepackage{epstopdf}
\usepackage{amssymb}
\usepackage{amsfonts}
\usepackage{amsmath}
\usepackage{bm}

\usepackage{multicol}        
\usepackage[bottom]{footmisc}

\makeindex             

\begin{document}

\title*{Condensed matter and AdS/CFT}
\author{Subir Sachdev}
\institute{Subir Sachdev \at Department of Physics, Harvard University, Cambridge MA 02138,\\ \email{sachdev@physics.harvard.edu}}
%
%
\maketitle

\textsl{\noindent
Lectures at the 5th Aegean summer school, ``From gravity to thermal gauge theories: the AdS/CFT correspondence'', Adamas, Milos Island, Greece, September 21-26, 2009,\\ and\\
the De Sitter Lecture Series in Theoretical Physics 2009, University of Groningen, November 16-20, 2009.}
~\\
~\\~\\~\\

\abstract{I review two classes of strong coupling problems in condensed matter
physics, and describe insights gained by application of the AdS/CFT correspondence.
The first class concerns non-zero temperature dynamics and transport in the vicinity
of quantum critical points described by relativistic field theories. I describe how relativistic structures
arise in models of physical interest, present results for their quantum critical crossover functions and magneto-thermoelectric hydrodynamics. The second class concerns symmetry breaking transitions
of two-dimensional systems 
in the presence of gapless electronic excitations at isolated points or along lines ({\em i.e.} Fermi surfaces)
in the Brillouin zone. I describe the scaling structure of a recent theory of the Ising-nematic
transition in metals, and discuss its possible connection to theories of Fermi surfaces obtained from
simple AdS duals.
}

\section{Introduction}
\label{sec:ssintro}

The past couple of decades have seen vigorous theoretical activity on the quantum phases
and phase transitions
of correlated electron systems in two spatial dimensions. Much of this work has been motivated
by the cuprate superconductors, but the list of interesting materials continues to increase unabated \cite{sssolvay}.

Methods from field theory have had a strong impact on much of this work. Indeed, they have become
part of the standard toolkit of condensed matter physicists. In these lectures, I focus on two classes of strong-coupling
problems which have not yielded accurate solutions via the usual arsenal of field-theoretic methods. 
I will also discuss how the AdS/CFT correspondence, discovered by string theorists, has already allowed
substantial progress on some of these problems, and offers encouraging prospects for future progress.

The first class of strong-coupling problems are associated with the real-time, finite temperature behavior
of strongly interacting quantum systems, especially those near quantum critical points. Field-theoretic
or numerical methods often allow accurate determination of the {\em zero\/} temperature or
of {\em imaginary time} correlations at non-zero temperatures. However, these methods usual fail
in the real-time domain at non-zero temperatures, particularly at times greater than $\hbar/k_ BT$, where $T$ is the 
absolute temperature. In systems near quantum critical points the natural scale for correlations is $\hbar/k_ BT$ itself,
and so lowering the temperature in a numerical study does not improve the situation.

The second class of strong-coupling problems arise near two-dimensional quantum critical points with fermionic excitations.
When the fermions have a massless Dirac spectrum, with zero excitation energy at a finite number of points in the Brillouin
zone, conventional field-theoretic methods do allow significant progress. However, in metallic systems, the fermionic
excitations have zeros along a line in the Brillouin zone (the Fermi surface), 
allowing a plethora of different low energy modes.
Metallic quantum critical points play a central role in many experimental systems, but the interplay between
the critical modes and the Fermi surface has not been fully understood (even at zero temperarture).
Readers interested only in this second class of problems can jump ahead to Section~\ref{sec:ssdwave}.

These lectures will start with a focus on the first class of strong-coupling problems.
We will begin in Section~\ref{sec:ssmodel} by introducing a variety of model systems and their
quantum critical points; these are motivated by recent experimental and theoretical developments.
We will use these systems to introduce basic ideas on the finite temperature crossovers near
quantum critical points in Section~\ref{sec:sscross}. In Section~\ref{sec:sstrans}, we will focus on the important {\em quantum critical
region\/} and present a general discussion of its transport properties.
An important recent development has been the complete exact solution, via the AdS/CFT correspondence, 
of the dynamic and transport
properties in the quantum critical region of a variety of (supersymmetric) 
model systems in two and higher dimensions: this
will be described in Section~\ref{sec:ssexact}. The exact solutions are found to agree with the earlier
general ideas discussed here in Section~\ref{sec:sstrans}.
As has often been the case in the history of physics, the existence of a new class of solvable models leads to new
and general insights which apply to a much wider class of systems, almost all 
of which are not exactly solvable. This has also been the case here, as we will review in Section~\ref{sec:sshydro}:
a hydrodynamic theory of the low frequency transport properties has been developed, and has led
to new relations between a variety of thermo-electric transport co-efficients. 

The latter part of these 
lectures will turn to the second class of strong coupling problems, by describing the role
of fermions near quantum critical points. In Section~\ref{sec:ssdwave} we will consider some simple symmetry
breaking transitions in $d$-wave superconductors. Such superconductors have fermionic excitations with a massless Dirac
spectrum, and we will show how they become critical near the quantum phase transition. We will review how
the field-theoretic $1/N$ expansion does allow solution of a large class of such problems. Finally, in Section~\ref{sec:ssmetal}
we will consider phase transitions of metallic systems with Fermi surfaces. We will discuss how the $1/N$ expansion
fails here, and review the results of recent work involving the AdS/CFT correspondence. 

Some portions of the discussions below have been adapted from
other review articles by the author \cite{altenberg,carr}.

\section{Model systems and their critical theories}
\label{sec:ssmodel}

\subsection{Coupled dimer antiferromagnets} 
\label{sec:sslgw}

\begin{figure}
\centering
 \includegraphics[width=\linewidth]{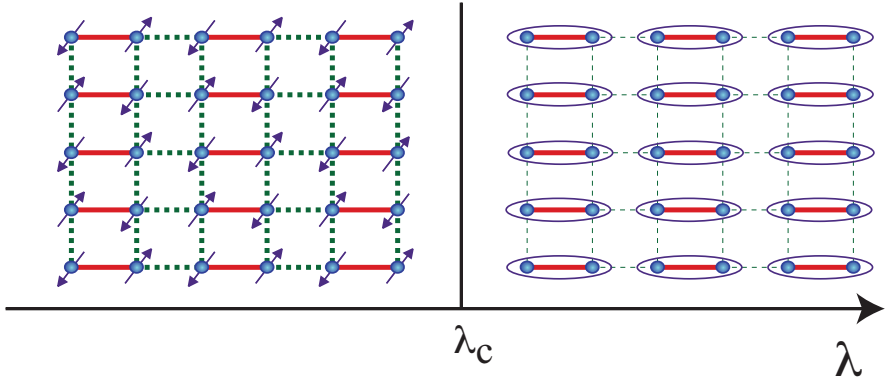}
 \caption{The coupled dimer antiferromagnet. The full red lines represent an exchange interaction $J$, while the dashed green lines have exchange $J/\lambda$. The ellispes represent a singlet valence
 bond of spins $(|\uparrow \downarrow \rangle - | \downarrow \uparrow \rangle )/\sqrt{2}$.}
\label{fig:ssdimer}
\end{figure}

Some of the best studied examples of quantum phase transitions arise in insulators with unpaired $S=1/2$ electronic
spins residing on the sites, $i$, of a regular lattice. Using $S^a_i$ ($a=x,y,z$) to represent the spin $S=1/2$ operator
on site $i$, the low energy spin excitations are described by the Heisenberg exchange Hamiltonian
\begin{equation}
H_J = \sum_{i<j} J_{ij} S^a_i \cdot S^a_j + \ldots
\label{eq:ssHJ}
\end{equation}
where $J_{ij} > 0$ is the antiferromagnetic exchange interaction. We will begin with a simple realization of 
this model is illustrated in Fig.~\ref{fig:ssdimer}. 
The $S=1/2$ spins reside on the sites of a square lattice, and have nearest neighbor exchange equal
to either $J$ or $J/\lambda$. Here $\lambda \geq 1$ is a tuning parameter which induces a quantum phase
transition in the ground state of this model. 

At $\lambda = 1$, the model has full square lattice symmetry,
and this case is known to have a N\'eel ground state which breaks spin rotation symmetry. This state has a
checkerboard polarization of the spins, just as found in the classical ground state, and as illustrated 
on the left side of Fig.~\ref{fig:ssdimer}. It can be characterized by a vector order parameter $\varphi^a$
which measures the staggered spin polarization
\begin{equation}
\varphi^a = \eta_i S^a_i
\end{equation}
where $\eta_i=\pm 1$ on the two sublattices of the square lattice. In the N\'eel state we have $\langle \varphi^a \rangle \neq  0$,
and we expect that the low energy excitations can be described by long wavelength fluctuations of a field $\varphi^a (x, \tau)$ over
space, $x$, and imaginary time $\tau$.

On the other hand, for $\lambda \gg 1$ it is evident from Fig.~\ref{fig:ssdimer} that the ground state preserves 
all symmetries of the Hamiltonian: it has total spin $S=0$ and can be considered to be a product of nearest
neighbor singlet valence bonds on the $J$ links. It is clear that this state cannot be smoothly connected
to the N\'eel state, and so there must at least one quantum phase transition as a function $\lambda$. 

Extensive quantum Monte Carlo simulations \cite{sstroyer,ssmatsu,ssjanke} 
on this model have shown there is a direct phase
transition between these states at a critical $\lambda_c$, as in Fig.~\ref{fig:ssdimer}. 
The value of $\lambda_c$
is known accurately, as are the critical exponents characterizing a second-order quantum phase
transition. These critical exponents are in excellent agreement with the simplest proposal for the critical
field theory, \cite{ssjanke} which can be obtained via conventional Landau-Ginzburg arguments. Given the vector
order parameter $\varphi^a$, we write down the action in $d$ spatial and one time dimension,
\begin{equation}
\mathcal{S}_{LG} = \int d^d r d\tau \left[ \frac{1}{2} \left[ (\partial_\tau \varphi^a )^2  + v^2 ( \nabla \varphi^a )^2 + s ( \varphi^a)^2 \right]
+ \frac{u}{4} \left[ (\varphi^a)^2 \right]^2 \right], \label{eq:ssslg}
\end{equation}
as the simplest action expanded in gradients and powers of $\varphi^a$ which is consistent will all
the symmetries of the lattice antiferromagnet.
The transition is now tuned by varying $s \sim (\lambda - \lambda_c)$. Notice that this model 
is identical to the Landau-Ginzburg theory for the thermal phase transition in a $d+1$ dimensional ferromagnet,
because time appears as just another dimension. As an example of the agreement: the critical exponent of the correlation
length, $\nu$, has the same value, $\nu = 0.711 \ldots$, to three significant digits in a quantum Monte Carlo study of the coupled
dimer antiferromagnet,\cite{ssjanke} and in a 5-loop analysis \cite{ssvicari} of the renormalization group fixed point of $\mathcal{S}_{LG}$
in $d=2$. 
Similar excellent agreement is obtained for the double-layer antiferromagnet \cite{sssandsca,ssmatsushita}
and the coupled-plaquette antiferromagnet.\cite{ssafa}

In experiments, the best studied realization of the coupled-dimer antiferromagnet is TlCuCl$_3$. In this crystal, the dimers are coupled
in all three spatial dimensions, and the transition from the dimerized state to the N\'eel state can be induced by application of pressure.
Neutron scattering experiments by Ruegg and collaborators \cite{ssruegg} have 
clearly observed the transformation in the excitation spectrum across the transition,
and these observations are in good quantitative agreement with theory\cite{sssolvay}.

\subsection{Deconfined criticality}
\label{sec:ssdeconfine}

We now consider an analog of transition discussed in Section~\ref{sec:sslgw}, but for a Hamiltonian $H=H_0 + \lambda H_1$ which has
full square lattice symmetry at all $\lambda$. For $H_0$, we choose a form of $H_J$, with $J_{ij}= J$ for all nearest
neighbor links. Thus at $\lambda=0$ the ground state has N\'eel order, as in the left panel of Fig.~\ref{fig:ssdimer}.
We now want to choose $H_1$ so that increasing $\lambda$ leads to a spin singlet state with spin rotation symmetry restored.
A large number of choices have been made in the literature, and the resulting ground state invariably \cite{ssrsl} has valence bond solid
(VBS) order; a VBS state has been observed in the organic antiferromagnet EtMe$_3$P[Pd(dmit)$_2$]$_2$ \cite{sskato1,sskato2}.  
The VBS state is superficially similar to the dimer singlet state in the right panel of Fig.~\ref{fig:ssdimer}:
the spins primarily form valence bonds with near-neighbor sites. However, because of the square lattice symmetry of the Hamiltonian, a columnar arrangement of the valence bonds as in Fig.~\ref{fig:ssdimer}, breaks the square lattice rotation
symmetry; there are 4 equivalent columnar states, with the valence bond columns running along different directions. 
More generally, a VBS state is a spin singlet state, with a non-zero degeneracy due to a spontaneously broken lattice
symmetry. Thus a direct transition between the N\'eel and VBS states involves two distinct broken symmetries:
spin rotation symmetry, which is broken only in the N\'eel state, and a lattice rotation symmetry, which is broken only
in the VBS state. The rules of Landau-Ginzburg-Wilson theory imply that there can be no generic second-order
transition between such states.

It has been argued that a second-order N\'eel-VBS transition can indeed occur \cite{sssenthil}, but the critical theory is not expressed
directly in terms of either order parameter. It involves a fractionalized bosonic spinor $z_\alpha$ ($\alpha = \uparrow,
\downarrow$), and an emergent gauge field $A_\mu$. 
The key step is to express the vector field $\varphi^a$ in terms of $z_\alpha$ by
\begin{equation}
\varphi^a = z_\alpha^\ast {\sigma}^a_{\alpha \beta} z_\beta
\label{eq:ssPhiz}
\end{equation}
where ${\sigma}^a$ are the $2\times2$ Pauli matrices. Note that this mapping from $\varphi^a$ to $z_\alpha$
is redundant. We can make a spacetime-dependent change in the phase of the $z_\alpha$ by the field $\theta(x,\tau)$
\begin{equation}
z_\alpha \rightarrow e^{i \theta} z_\alpha
\label{eq:ssgauge}
\end{equation}
and leave $\varphi^a$ unchanged. All physical properties must therefore also be invariant under Eq.~(\ref{eq:ssgauge}),
and so the quantum field theory for $z_\alpha$ has a U(1) gauge invariance, much like that found in quantum electrodynamics.
The effective action for the $z_\alpha$ therefore requires introduction of an `emergent' 
U(1) gauge field $A_\mu$ (where $\mu = x, \tau$ is a 
three-component spacetime index). The field $A_\mu$ is unrelated the electromagnetic field, but is an internal
field which conveniently describes the couplings between the spin excitations of the antiferromagnet.  
As we did for $\mathcal{S}_{LG}$,
we can write down the quantum field theory for $z_\alpha$ and $A_\mu$ by the constraints of symmetry and gauge invariance,
which now yields
\begin{equation}
\mathcal{S}_z =  \int d^2 r d \tau \biggl[
|(\partial_\mu -
i A_{\mu}) z_\alpha |^2 + s |z_\alpha |^2  + u (|z_\alpha |^2)^2 + \frac{1}{2g^2}
(\epsilon_{\mu\nu\lambda}
\partial_\nu A_\lambda )^2 \biggl] \label{eq:ssSz}
\end{equation}
For brevity, we have now used a ``relativistically'' invariant notation, and scaled away the spin-wave velocity $v$; the values
of the couplings $s,u$ are different from, but related to, those in $\mathcal{S}_{LG}$. The Maxwell action for $A_\mu$ is generated from 
short distance $z_\alpha$ fluctuations, and it makes $A_\mu$ a dynamical field; its coupling $g$ is unrelated
to the electron charge. 
The action $\mathcal{S}_z$ is a valid description of the N\'eel state for $s<0$ (the critical upper value of $s$ will have fluctuation
corrections away from 0), where the gauge theory enters a Higgs phase with $\langle z_\alpha \rangle \neq 0$. This description of the N\'eel state as a Higgs phase has an analogy with the Weinberg-Salam theory of the weak interactions---in the latter case it is hypothesized that the condensation of a Higgs boson gives a mass to the $W$ and $Z$ gauge bosons, whereas here the condensation of $z_\alpha$ quenches the $A_\mu$ gauge boson.
As written, the $s>0$ phase of $\mathcal{S}_z$ is a `spin liquid' state with a $S=0$ collective gapless excitation associated with the 
$A_\mu$ photon. Non-perturbative effects \cite{ssrsl} associated with the monopoles in $A_\mu$ (not discussed here), show that this spin liquid is
ultimately unstable to the appearance of VBS order. 

Numerical studies of the N\'eel-VBS transition have focussed on a specific lattice antiferromagnet proposed by 
Sandvik \cite{sssandvik,sssandvik2,ssmelkokaul}. There is strong
evidence for VBS order proximate to the N\'eel state, along with persuasive evidence of a second-order transition.
However, some studies \cite{sswiese,sskuklov} support a very weak first order transition.

\subsection{Graphene}
\label{sec:ssgraphene}

The last few years have seen an explosion in experimental and theoretical studies \cite{ssneto} of graphene: a single hexagonal layer of carbon atoms.
At the currently observed temperatures, there is no evident broken symmetry in the electronic excitations, and so it is not
conventional to think of graphene as being in the vicinity of a quantum critical point. However, graphene does indeed undergo
a bona fide quantum phase transition, but one without any order parameters or broken symmetry. This transition may be viewed as 
being `topological' in character, and is associated with a change in nature of the Fermi surface as a function of carrier density.

\begin{figure}
\centering
 \includegraphics[width=3in]{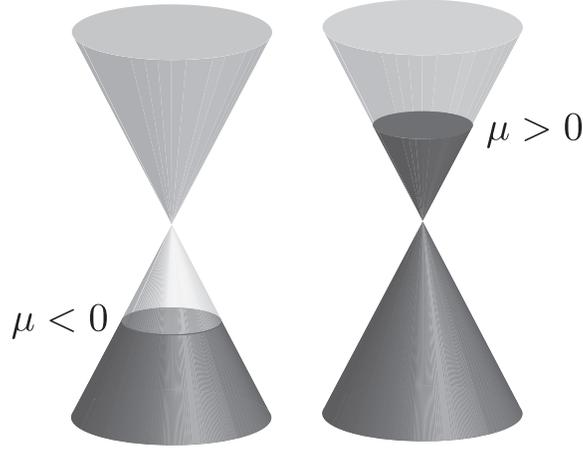}
 \caption{Dirac dispersion spectrum for graphene showing a `topological' quantum phase transition from a hole Fermi surface for $\mu<0$ to a electron Fermi surface
 for $\mu > 0$.}
\label{fig:ssgraphene}
\end{figure}

Pure, undoped graphene has a conical electronic dispersion spectrum at two points in the Brillouin zone, with the Fermi energy at the particle-hole
symmetric point at the apex of the cone. So there is no Fermi surface, just a Fermi point, where the electronic energy vanishes, and pure graphene
is a `semi-metal'. By applying a gate voltage, the Fermi energy can move away from this symmetric point, and a circular Fermi surface develops,
as illustrated in Fig.~\ref{fig:ssgraphene}. 
The Fermi surface is electron-like for one sign of the bias, and hole-like for the other sign.
This change from electron to hole character as a function of gate voltage constitutes the quantum phase transition in graphene.
As we will see below, with regard to its dynamic properties near zero bias, graphene behaves in almost all respects like a canonical quantum critical system.

The field theory for graphene involves fermionic degrees of freedom. Representing the electronic orbitals near one of the Dirac points by
the two-component fermionic spinor $\Psi_s$, where $s$ is a sublattice index (we suppress spin and `valley' indices), we have the effective
electronic action
\begin{eqnarray}
\mathcal{S}_\Psi &=& \int d^2 r \int d \tau\,  \Psi^\dagger_s \left[ (\partial_\tau + 
i A_\tau - \mu) \delta_{ss'} + i v_F \tau^x_{ss'} \partial_x
+ i v_F \tau^y_{ss'} \partial_y \right] \Psi_{s'} \nonumber \\
&~&~~~~~~~~~~~~+ \frac{1}{2g^2} \int \frac{d^2 q}{4 \pi^2} \int d\tau\, \frac{q}{2\pi} \left| A_\tau ({\bf q},\tau) \right|^2, 
\label{eq:ssgraph}
\end{eqnarray}
where $\tau^i_{ss'}$ are Pauli matrices in the sublattice space, $\mu$ is the chemical potential, $v_F$ is the Fermi velocity,
and $A_\tau$ is the scalar potential mediating the Coulomb interaction
with coupling $g^2 = e^2/\epsilon$ ($\epsilon$ is a dielectric constant). 
This theory undergoes a quantum phase transition as a function of $\mu$, at $\mu=0$,
similar in many ways to that of $\mathcal{S}_{LG}$ as a function of $s$. The interaction between the fermionic excitations here
has coupling $g^2$, which is the analog of the non-linearity $u$ in $\mathcal{S}_{LG}$. The strength of the interactions is
determined by the dimensionless `fine structure constant'
$\alpha = g^2/(\hbar v_F)$ which is of order unity in graphene.
While $u$ flows to a
non-zero fixed point value under the renormalization group, $\alpha$ flows logarithmically slowly to zero. For many purposes, it is safe
to ignore this flow, and to set $\alpha$ equal to a fixed value. 

\section{Finite temperature crossovers}
\label{sec:sscross}
The previous section has described four model systems at $T=0$: we examined the change in the nature of the ground
state as a function of some tuning parameter, and motivated a quantum field theory which describes the low energy excitations
on both sides of the quantum critical point.

\begin{figure}
\centering
 \includegraphics[width=3.9in]{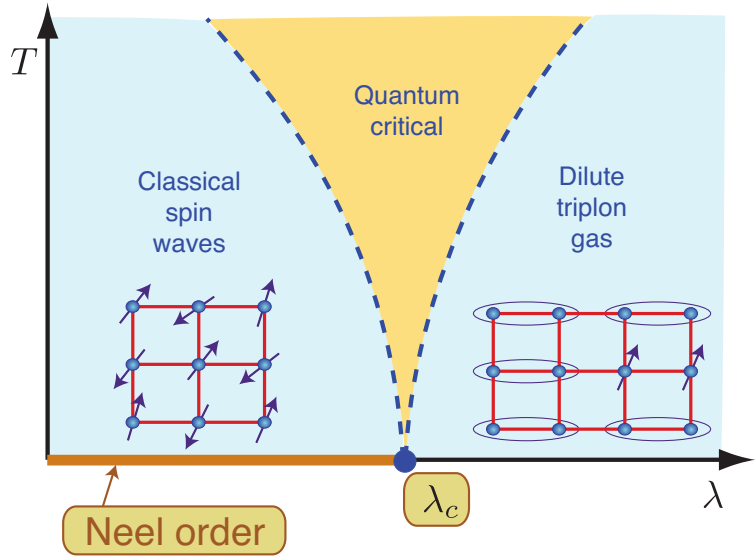}
 \caption{Finite temperature crossovers of the coupled dimer antiferromagnet in Fig.~\ref{fig:ssdimer}.}
\label{fig:sscross_dimer}
\end{figure}

We now turn to the important question of the physics at non-zero temperatures. All of the models share some common features,
which we will first explore for the coupled dimer antiferromagnet.  For $\lambda > \lambda_c$ (or $s>0$ in $\mathcal{S}_{LG}$),
the excitations consist of a triplet of $S=1$ particles (the `triplons'), which can be understood perturbatively in the large
$\lambda$ expansion as an excited $S=1$ state on a dimer, hopping between dimers (see Fig.~\ref{fig:sscross_dimer}).
The mean field theory tells us that the excitation energy of this dimer vanishes as $\sqrt{s}$ upon approaching
the quantum critical point. Fluctuations beyond mean field, described by $\mathcal{S}_{LG}$, show that the exponent
is modified to $s^{z \nu}$, where $z=1$ is the dynamic critical exponent, and $\nu$ is the correlation length exponent.
Now imagine turning on a non-zero temperature. As long as $T$ is smaller than the triplon gap, {\em i.e.\/}
$T < s^{z \nu}$, we expect a description in terms of a dilute gas of thermally excited triplon particles. This leads to the behavior
shown on the right-hand-side of Fig.~\ref{fig:sscross_dimer}, delimited by the crossover indicted by the dashed line.
Note that the crossover line approaches $T=0$ only at the quantum critical point.

Now let us look a the complementary behavior at $T>0$ on the N\'eel-ordered side of the transition, with $s<0$. 
In two spatial dimensions, 
thermal fluctuations prohibit the breaking of a non-Abelian symmetry at all $T>0$, and so spin rotation symmetry
is immediately restored. Nevertheless, there is an exponentially large spin correlation length, $\xi$, and at distances shorter
than $\xi$ we can use the ordered ground state to understand the nature of the excitations. Along with the spin-waves, we also
found the longitudinal `Higgs' mode with energy $\sqrt{-2s}$ in mean field theory. Thus, just as was this case for $s>0$, we
expect this spin-wave+Higgs picture to apply at all temperatures lower than the natural energy scale; {\em i.e.\/} 
for $T<(-s)^{z\nu}$. This leads to the crossover boundary shown on the left-hand-side
of Fig.~\ref{fig:sscross_dimer}. 

Having delineated the physics on the two sides of the transition, we are left with the crucial {\em quantum critical\/} region in the
center of Fig.~\ref{fig:sscross_dimer}. This is present for $T > |s|^{z \nu}$, {\em i.e.\/} at {\em higher\/} temperatures in the
vicinity of the quantum critical point. To the left of the quantum critical region, we have a description of the dynamics and transport
in terms of an effectively classical model of spin waves: this is the `renormalized classical' regime of Ref.~\cite{sschn}.
To the right of the quantum critical region, we again have a regime of classical dynamics, but now in terms of a Boltzmann
equation for the triplon particles. A key property of quantum critical region is that there is no description in terms
of either classical particles or classical waves at the times of order the typical relaxation time, $\tau_r$, of thermal excitations.
Instead, quantum and thermal effects are equally important, and involve the non-trivial dynamics of the fixed-point
theory describing the quantum critical point. Note that while the fixed-point theory applies only at a single point ($\lambda=\lambda_c$)
at $T=0$, its influence broadens into the quantum critical region at $T>0$. Because there is no characteristing energy scale
associated with the fixed-point theory, $k_B T$ is the only energy scale available to determine $\tau_r$ at non-zero
temperatures. Thus, in the quantum critical region \cite{sscsy}
\begin{equation}
\tau_r = \mathcal{C}\frac{\hbar}{k_B T}
\label{eq:ssrelax}
\end{equation}
where $\mathcal{C}$ is a universal constant dependent only upon the universality class of the fixed point theory {\em i.e.\/}
it is universal number just like the critical exponents. This value of $\tau_r$ determines the `friction coefficients' associated
with the dissipative relaxation of spin fluctuations in the quantum critical region. It is also important for the transport
co-efficients associated with conserved quantities, and this will be discussed in Section~\ref{sec:sstrans}.

Let us now consider the similar $T>0$ crossovers for the other models of Section~\ref{sec:ssmodel}.

The N\'eel-VBS transition of Section~\ref{sec:ssdeconfine}
has crossovers very similar to those in Fig.~\ref{fig:sscross_dimer}, with one important
difference. The VBS state breaks a discrete lattice symmetry, and this symmetry remains broken for a finite range
of non-zero temperatures. Thus, within the right-hand 'triplon gas' regime of Fig.~\ref{fig:sscross_dimer}, there is a 
phase transition line at a critical temperature $T_{\rm VBS}$. The value of $T_{\rm VBS}$ vanishes very rapidly
as $s \searrow 0$, and is controlled by the non-perturbative monopole effects which were briefly noted
in Section~\ref{sec:ssdeconfine}.

\begin{figure}
\centering
 \includegraphics[width=3.5in]{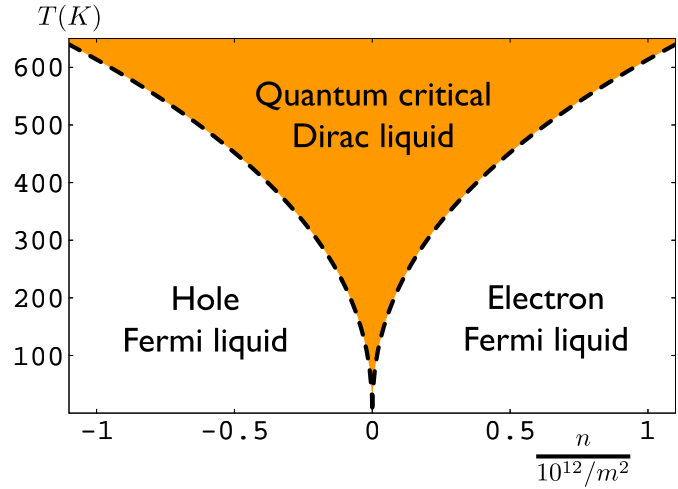}
 \caption{Finite temperature crossovers of graphene as a function of electron density $n$ (which is tuned by $\mu$ in
 Eq.~(\ref{eq:ssgraph})) and temperature, $T$. Adapted from Ref.~\cite{sssheehy}.}
\label{fig:sscross_graphene}
\end{figure}
For graphene, the discussion above applied to Fig.~\ref{fig:ssgraphene} leads to the crossover diagram
shown in Fig.~\ref{fig:sscross_graphene}, as noted by Sheehy and Schmalian \cite{sssheehy}.
We have the Fermi liquid regimes of the electron- and hole-like Fermi surfaces on either side of the critical point,
along with an intermediate quantum critical Dirac liquid. A new feature here is related to the logarithmic flow of the 
dimensionless `fine structure constant' $\alpha$ controlling the Coulomb interactions, which was noted in Section~\ref{sec:ssgraphene}.
In the quantum critical region, this constant takes the typical value $\alpha \sim 1/\ln (1/T)$. Consequently for the relaxation
time in Eq.~(\ref{eq:ssrelax}) we have $\mathcal{C} \sim \ln^2 (1/T)$. This time determines both the width of the
electron spectral functions, and also the transport co-efficients, as we will see in Section~\ref{sec:sstrans}.

\section{Quantum critical transport}
\label{sec:sstrans}

We now turn to the `transport' properties in the quantum critical region: we consider the response
functions associated with any globally conserved quantity.
For the antiferromagnetic systems in Sections~\ref{sec:sslgw} and~\ref{sec:ssdeconfine},
this requires consideration of the transport of total spin, and the associated spin conductivities and diffusivities.
For graphene, we can consider charge and momentum transport. Our discussion below will also apply to the 
superfluid-insulator transition: for bosons in a periodic potential, this transition is described \cite{ssfwgf} by a field theory
closely related to that in Eq.~(\ref{eq:ssslg}). However, we will primarily use a language appropriate to charge transport
in graphene below. We will describe the properties of a generic strongly-coupled quantum critical point and mention, where
appropriate,
the changes due to the logarithmic flow of the coupling in graphene.

In traditional condensed matter physics, transport is described by identifying the low-lying
excitations of the quantum ground state, and writing down `transport equations' for the conserved charges carried by them.
Often, these excitations have a particle-like nature, such as the `triplon' particles of Fig.~\ref{fig:sscross_dimer} or the electron
or hole quasiparticles of the Fermi liquids in Fig.~\ref{fig:sscross_graphene}. In other cases, the low-lying excitations are waves, such as the spin-waves in Fig.~\ref{fig:sscross_dimer}, and their transport is described by a non-linear wave equation (such as the 
Gross-Pitaevski equation). 
However, as we have discussed in Section~\ref{sec:sscross} neither description is 
possible in the quantum critical region, because the excitations do not have a particle-like
or wave-like character. 

Despite the absence of an intuitive description of the quantum critical dynamics, we can expect that the transport
properties should have a universal character determined by the quantum field theory of the quantum critical point. In addition
to describing single excitations, this field theory also determines the $S$-matrix of these excitations by the renormalization
group fixed-point value of the couplings, and these should be sufficient to determine transport properties \cite{ssdamle}. 
The transport co-efficients, and the relaxation time to local
equilibrium, are not proportional to a mean free scattering time between the excitations, as is
the case in the Boltzmann theory of quasiparticles. Such a time would typically depend upon the interaction
strength between the particles. Rather, the system behaves like a ``perfect fluid'' in which the relaxation
time is as short as possible, and is determined universally by the absolute temperature, as indicated in Eq.~(\ref{eq:ssrelax}).
 Indeed, it was
conjectured in Ref.~\cite{ssbook} that the relaxation time in Eq.~(\ref{eq:ssrelax}) is a generic lower bound
for interacting quantum systems. Thus the non-quantum-critical regimes of all the phase diagrams in Section~\ref{sec:sscross}
have relaxation times which are all longer than Eq.~(\ref{eq:ssrelax}).

The transport co-efficients of this quantum-critical perfect fluid also do not depend upon the interaction
strength, and can be connected to the fundamental constants of nature. In particular, the electrical conductivity, $\sigma$,
is given by (in two spatial dimensions) \cite{ssdamle}
\begin{equation}
\sigma_Q = \frac{e^{\ast 2}}{h} \Phi_\sigma, \label{ssds}
\end{equation}
where $\Phi_\sigma$ is a universal dimensionless constant of order unity, and we have added the subscript $Q$ to emphasize that this is the conductivity for the case of graphene with the Fermi level at the Dirac point (for the superfluid-insulator
transition, this would correspond to bosons at integer filling) with
no impurity scattering, and at zero magnetic field. Here $e^\ast$ is the charge of the carriers: for a superfluid-insulator
transition of Cooper pairs, we have $e^\ast = 2e$, while for graphene we have $e^\ast = e$.
The renormalization group flow of the `fine structure constant' $\alpha$ of graphene to zero at asymptotically low $T$, allows
an exact computation in this case \cite{ssmarkus}: $\Phi_\sigma \approx 0.05 \ln^2 (1/T)$. For the superfluid-insulator
transition, $\Phi_\sigma$ is $T$-independent (this is the generic situation with non-zero fixed point values of the interaction \cite{sslongtime}) but it has only been computed \cite{ssbook,ssdamle} to 
leading order in expansions
in $1/N$ (where $N$ is the number of order parameter components) and in $3-d$ (where $d$ is the spatial dimensionality).
However, both expansions are neither straightforward nor rigorous, and 
require a physically motivated resummation of the
bare perturbative expansion to all orders. It would therefore be valuable to have exact solutions of quantum critical
transport where the above results can be tested, and we turn to such solutions in the next section.

In addition to charge transport, we can also consider momentum transport. This was considered in the context of applications
to the quark-gluon plasma \cite{sskss}; application of the analysis of Ref.~\cite{ssdamle} shows that the viscosity, $\eta$, is given by
\begin{equation}
\frac{\eta}{s} = \frac{\hbar}{k_B} \Phi_\eta, \label{sseta}
\end{equation}
where $s$ is the entropy density, and again $\Phi_\eta$ is a universal constant of order unity. 
The value of $\Phi_\eta$ has recently been computed \cite{ssgrapheneperfect} for graphene, and again has a logarithmic 
$T$ dependence because of the marginally irrelevant interaction: $\Phi_\eta \approx 0.008 \ln^2 (1/T)$.

We conclude this section by discussing  
some subtle aspects of the physics behind the seemingly simple result quantum-critical in Eq.~(\ref{ssds}).
For simplicity, we will consider the case of a ``relativistically'' invariant quantum critical point in 2+1 dimensions
(such as the field theories of Section~\ref{sec:sslgw} and~\ref{sec:ssdeconfine}, but marginally violated by graphene, a subtlety we ignore below).
Consider the retarded
correlation function of the charge density, $\chi (k, \omega)$, where $k = |{\bf k}|$
is the wavevector, and $\omega$ is frequency; the dynamic conductivity, $\sigma(\omega)$, is related to $\chi$ by the 
Kubo formula, 
\begin{equation}
\sigma (\omega) = \lim_{k \rightarrow 0} \frac{-i \omega}{k^2} \chi (k,
\omega).
\label{sskubo}
\end{equation}
It was argued in Ref.~\cite{ssdamle} that 
despite the absence of particle-like excitations of the critical ground state, the central characteristic of the transport
is a crossover from collisionless to collision-dominated transport. At high frequencies or low temperatures,
the limiting form for $\chi$ reduces to that at $T=0$, which is completely determined by relativistic and scale invariance 
and current conversion upto an overall constant
\begin{equation}
\chi (k, \omega)    = \frac{e^{\ast 2}}{h} K \frac{k^2}{\sqrt{v^2 k^2 - (\omega+i \eta)^2}}~~,~~\sigma(\omega) = 
\frac{e^{\ast 2}}{h} K ~~;~~~\mbox{$\hbar\omega \gg k_B T$,} 
\label{ssd2c}
\end{equation}
where $K$ is a universal number \cite{ssmpaf}.
However,  phase-randomizing collisions are intrinsically present in any strongly interacting critical point (above one
spatial dimension)
and these lead to relaxation of perturbations to local equilibrium and the consequent emergence
of hydrodynamic behavior. So at low frequencies, we have instead an Einstein relation which determines the conductivity with
\begin{equation}
\chi (k, \omega)  =  e^{\ast 2} \chi_c \frac{Dk^2}{Dk^2 - i \omega}~~,~~\sigma(\omega) = e^{\ast 2} \chi_c D = 
\frac{e^{\ast 2}}{h} \Theta_1\Theta_2~~;~~~\mbox{$\hbar\omega \ll k_B T$,} 
\label{ssd2h}
\end{equation}
where $\chi_c$ is the compressibility and $D$ is the charge diffusion constant. Quantum critical scaling arguments show that
the latter quantities obey
\begin{equation}
\chi_c = \Theta_1 \frac{k_B T}{h^2 v^2}~~,~~ D = \Theta_2 \frac{h v^2}{k_B T}, 
\end{equation}
where $\Theta_{1,2}$ are universal numbers.
A large number of papers in the literature, 
particularly those on critical points in quantum Hall systems, have used the collisionless method of Eq.~(\ref{ssd2c}) to 
compute the conductivity.
However, the correct d.c. limit is given by Eq.~(\ref{ssd2h}),
and the universal constant in Eq.~(\ref{ssds}) is given by $\Phi_{\sigma} = \Theta_1 \Theta_2$.
Given the distinct physical interpretation of the collisionless and collision-dominated regimes,
we expect that $K \neq \Theta_1 \Theta_2$.
This has been shown in a resummed perturbation expansion for a number of 
quantum critical points \cite{ssbook}.

\section{Exact results for quantum critical transport}
\label{sec:ssexact}

The results of Section~\ref{sec:sstrans} were obtained by using physical arguments
to motivate resummations of perturbative expansions. Here we shall support the ad hoc
assumptions behind these results by examining an exactly solvable model of quantum critical transport.

The solvable model may be viewed as a generalization of the gauge theory in Eq.~(\ref{eq:ssSz}) to the maximal
possible supersymmetry. In 2+1 dimensions, this is known as $\mathcal{N}=8$ supersymmetry. Such a  theory with
the U(1) gauge group is free, and so we consider the non-Abelian Yang-Millis theory with a SU($N$) gauge group. The resulting
supersymmetric Yang-Mills (SYM) theory has only one coupling constant, which is the analog of the electric charge $g$
in Eq.~(\ref{eq:ssSz}). The matter content is naturally more complicated than the complex scalar $z_\alpha$ in 
Eq.~(\ref{eq:ssSz}), and also involves relativistic Dirac fermions as in Eq.~(\ref{eq:ssgraph}). However all the terms in the action
for the matter fields are also uniquely fixed by the single coupling constant $g$. Under the renormalization group, it is believed
that $g$ flows to an attractive fixed point at a non-zero coupling $g=g^\ast$; the fixed point then defines
a supersymmetric conformal field theory in 2+1 dimensions (a SCFT3), and we are interested here in the transport
properties of this SCFT3.

A remarkable recent advance has been the exact solution of this SCFT3 in the $N\rightarrow \infty$ limit
using the AdS/CFT correspondence \cite{ssimsy}.
The solution proceeds by a dual formulation as a four-dimensional supergravity theory on a spacetime with uniform negative 
curvature: anti-de Sitter space, or AdS$_4$. Remarkably, the solution is also easily extended to non-zero temperatures,
and allows direct computation of the correlators of conserved charges in real time. 
At $T>0$ a black hole appears in the gravity, resulting
in an AdS-Schwarzschild spacetime, and $T$ is also the Hawking temperature of the black hole; the real time solutions
also extend to $T>0$.

\begin{figure}
\centering
 \includegraphics[width=3.6in]{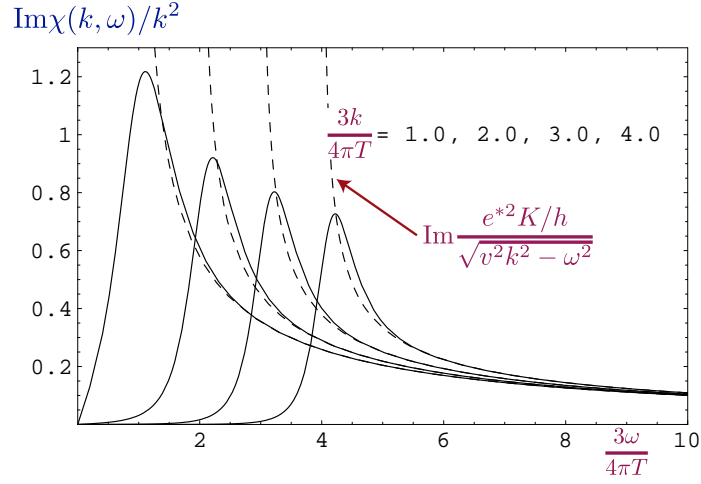}
 \caption{Spectral weight of the density correlation function of the SCFT3 with $\mathcal{N}=8$
 supersymmetry
 in the collisionless regime.}
\label{fig:sschi_collisionless}
\end{figure}

\begin{figure}
\centering
 \includegraphics[width=3.6in]{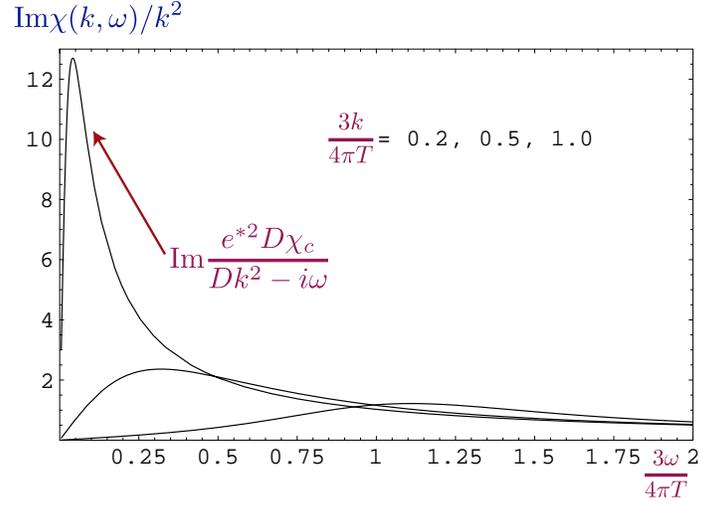}
 \caption{As in Fig.~\ref{fig:sschi_collisionless}, but for the collision-dominated regime.}
\label{fig:sschi_diff}
\end{figure}

The results of a full computation \cite{ssm2cft} of the density correlation function, $\chi (k, \omega)$ are shown
in Fig.~\ref{fig:sschi_collisionless} and~\ref{fig:sschi_diff}.
The most important feature of these results is that the expected limiting forms
in the collisionless (Eq.~(\ref{ssd2c})) and collision-dominated (Eq.~(\ref{ssd2h}) are obeyed.
Thus the results do display the collisionless to collision-dominated crossover at a frequency of order
$k_B T/\hbar$, as was postulated in Section~\ref{sec:sstrans}.

An additional important feature of the solution is apparent upon describing the full structure of 
both the density and current correlations. Using spacetime indices ($\mu,\nu=t,x,y$) we can represent these
as the tensor $\chi_{\mu\nu} ({\bf k}, \omega)$, where the previously considered $\chi \equiv \chi_{tt}$.
At $T>0$, we do not expect $\chi_{\mu\nu}$ to be relativistically covariant, and so can only constrain it by
spatial isotropy and density conservation. Introducing a spacetime momentum $p_\mu = (\omega, {\bf k})$,
and setting the velocity $v=1$,  these two constraints lead to the most general form
\begin{equation}
\chi_{\mu\nu} ({\bf k}, \omega) = \frac{e^{\ast 2}}{h} \sqrt{p^2} \Bigl( P^T_{\mu\nu}\, K^T (k,\omega)
  + P^L_{\mu\nu}\, K^L (k,\omega) \Bigr)
\label{eq:sscmn}
\end{equation}
where $p^2 = \eta^{\mu\nu} p_\mu p_\nu$ with $\eta_{\mu \nu} = \mbox{diag}(-1,1,1)$,
and $P^T_{\mu\nu}$ and $P^L_{\mu\nu}$ are
orthogonal projectors defined by
\begin{equation}
P^T_{00} = P^T_{0i} = P^T_{i0}=0~~,~~P^T_{ij} = \delta_{ij} -
\frac{k_i k_j}{k^2}~~,~~P^L_{\mu\nu} =
  \Big(\eta_{\mu\nu} - \frac{p_\mu p_\nu}{p^2}\Big) - P^T_{\mu\nu},
\end{equation}
with the indices $i,j$ running over the 2 spatial components. The two functions $K^{T,L} (k, \omega)$ define all the
correlators of the density and the current, and the results in Eqs.~(\ref{ssd2h}) and (\ref{ssd2c}) are obtained by
taking suitable limits of these functions. We will also need below the general identity 
\begin{equation}
K^T (0,\omega) = K^L (0,\omega),
\label{eq:sslt}
\end{equation}
which follows from the analyticity of the $T>0$ current correlations at ${\bf k } = 0$.

The relations of the previous paragraph are completely general and apply to any theory.
Specializing to the AdS-Schwarzschild solution of SYM3, the results were found to obey a simple
and remarkable identity \cite{ssm2cft}:
\begin{equation}
K^L (k, \omega) K^T (k, \omega) = \mathcal{K}^2
\label{eq:sssdual}
\end{equation}
where $\mathcal{K}$ is a known pure number, independent of $\omega$ and $k$. It was also shown that such a relation
applies to any theory which is equated to classical gravity on AdS$_4$, and is a consequence of the electromagnetic self-duality of its four-dimensional Maxwell sector. The combination of Eqs.~(\ref{eq:sslt})
and (\ref{eq:sssdual}) fully determines the $\chi_{\mu\nu}$ correlators at ${\bf k} = 0$: we find 
$K^L (0, \omega) = K^T (0, \omega) = \mathcal{K}$, from which it follows that the ${\bf k}=0$ conductivity
is frequency independent and that $\Phi_\sigma = \Theta_1 \Theta_2 = K= \mathcal{K}$. These last features are believed
to be special to theories which are equivalent to classical gravity, and not hold more generally.

We can obtain further insight into the interpretation of Eq.~(\ref{eq:sssdual}) by considering the field theory of the
superfluid-insulator transition of lattice bosons at integer filling. As we noted earliear, this is given by the 
field theory in Eq.~(\ref{eq:ssslg}) with the field $\varphi^a$ having 2 components. It is known that this 2-component
theory of relativistic bosons is equivalent to a dual relativistic theory, $\widetilde{\mathcal{S}}$ of vortices, under the well-known `particle-vortex'
duality \cite{ssdh}. Ref.~\cite{ssm2cft} considered the action of this particle-vortex duality on the correlation functions
in Eq.~(\ref{eq:sscmn}), and found the following interesting relations:
\begin{equation}
K^L (k, \omega) \widetilde{K}^T (k, \omega) = 1~~~,~~~K^T (k, \omega) \widetilde{K}^L (k, \omega) = 1
\label{eq:ssdual}
\end{equation}
where $\widetilde{K}^{L,T}$ determine the vortex current correlations in $\widetilde{\mathcal{S}}$ as in Eq.~(\ref{eq:sscmn}). 
Unlike Eq.~(\ref{eq:sssdual}),
Eq.~(\ref{eq:ssdual}) does {\em not} fully determine the correlation functions at ${\bf k} = 0$: it only serves
to reduce the 4 unknown functions $K^{L,T}$, $\widetilde{K}^{L,T}$ to 2 unknown functions. The key property here
is that while the theories $\mathcal{S}_{LG}$ and $\widetilde{\mathcal{S}}$ are dual to each other, they are not equivalent, and the theory $\mathcal{S}_{LG}$ is not self-dual.

We now see that Eq.~(\ref{eq:sssdual}) implies that the classical gravity theory of SYM3 is self-dual under
an analog of particle-vortex duality \cite{ssm2cft}. It is not expected that this self-duality will hold
when quantum gravity corrections are included; equivalently, the SYM3 at finite $N$ is expected
to have a frequency dependence in its conductivity at ${\bf k} = 0$. If we
apply the AdS/CFT correspondence to the superfluid-insulator transition, and approximate the latter theory
by classical gravity on AdS$_4$,  we immediately obtain the self-dual prediction for the conductivity, $\Phi_\sigma = 1$.
This value is not far from that observed in numerous experiments, and we propose here that the AdS/CFT correspondence
offers a rationale for understanding such observations.

\section{Hydrodynamic theory}
\label{sec:sshydro}

The successful comparison between the general considerations of Section~\ref{sec:sstrans}, and the exact solution
using the AdS/CFT correspondence in Section~\ref{sec:ssexact}, emboldens us to seek a more general theory of
low frequency ($\hbar \omega \ll k_B T$) transport in the quantum critical regime. We will again present our results
for the special case of a relativistic quantum critical point in 2+1 dimensions (a CFT3), but it is clear that similar considerations
apply to a wider class of systems. Thus we can envisage applications to the superfluid-insulator transition, and have presented scenarios under which
such a framework can be used to interpret measurements of the Nernst effect in the cuprates \cite{ssnernst}. 
We have also described a separate set of applications to graphene \cite{ssmarkus}: 
while graphene is strictly not a CFT3, the Dirac spectrum of electrons
leads to many similar results, especially in the inelastic collision-dominated regime associated with the quantum critical region.
These results on graphene are reviewed in a separate paper \cite{ssgraphrev}, where explicit microscopic computations are also discussed.

Our idea is to relax the restricted set of conditions under which the results of Section~\ref{sec:sstrans} were obtained.
We will work within the quantum critical regimes of the phase diagrams of Section~\ref{sec:sscross} but now allow
a variety of additional perturbations. First, we will move away from the particle-hole symmetric case, allow a finite density
of carriers. For graphene, this means that $\mu$ is no longer pinned at zero; for the antiferromagnets, we can apply
an external magnetic field; for the superfluid-insulator transition, the number density need not be commensurate
with the underlying lattice. For charged systems, such as the superfluid-insulator transition or graphene, we allow
application of an external magnetic field. Finally, we also allow a small density of impurities which can act as a sink
of the conserved total momentum of the CFT3. In all cases, the energy scale associated with these perturbations 
is assumed to be smaller than the dominant energy scale of the quantum critical region, which is $k_B T$.
The results presented below were obtained in two separate computations, associated with the methods described in 
Sections~\ref{sec:sstrans}
and~\ref{sec:ssexact}, and are described in the two subsections below.

\subsection{Relativistic magnetohydrodynamics}
\label{sec:ssmagneto}

With the picture of relaxation to local equilibrium at frequencies $\hbar \omega \ll k_B T$ developed in Ref.~\cite{ssdamle},
we postulate that the equations of relativistic magnetohydrodynamics should describe the low frequency transport.
The basic principles involved in such a hydrodynamic computation go back to the nineteenth century: conservation
of energy, momentum, and charge, and the constraint of the positivity of entropy production. Nevertheless,
the required results were not obtained until our recent work \cite{ssnernst}: the general case of a CFT3 in the presence of
a chemical potential, magnetic field, and small density of impurities is very intricate, and the guidance
provided by the dual gravity formulation was very helpful to us. In this approach, we do not have quantitative
knowledge of a few transport co-efficients, and this is complementary to our ignorance of the effective
couplings in the dual gravity theory to be discussed in Section~\ref{sec:ssdyon}.

The complete hydrodynamic analysis can be found in Ref.~\cite{ssnernst}. The analysis is intricate, but is mainly 
a straightforward adaption of the classic procedure outlined by Kadanoff and Martin \cite{sskm} to the relativistic field
theories which describe quantum critical points. We list the steps: 
\begin{enumerate}
\item Identify the conserved quantities, which are the energy-momentum tensor, $T^{\mu\nu}$, and the particle number current, $J^\mu$. 
\item Obtain the real time 
equations of motion, which express the conservation laws:
\begin{equation}
\partial_\nu T^{\mu\nu} = F^{\mu\nu}J_\nu~~~,~~~\partial_\mu J^\mu = 0;
\end{equation}
here $F^{\mu\nu}$ represents the externally applied electric and magnetic fields which can change the net
momentum or energy of the system, and we have not written a term describing momentum relaxation
by impurities. 
\item Identify the state variables which characterize the local thermodynamic state---we choose these to be the density, $\rho$, the temperature $T$, and an average velocity $u^\mu$. 
\item Express $T^{\mu\nu}$ and $J^\mu$ in terms of the state variables and their spatial and temporal
gradients; here we use the properties of the observables under a boost by the velocity $u^\mu$, and thermodynamic quantities like the energy density, $\varepsilon$, and the pressure,
$P$, which are determined from $T$ and $\rho$ by the equation of state of the CFT. We also introduce transport 
co-efficients associated with the gradient terms. 
\item Express the equations of motion in terms
of the state variables, and ensure that the entropy production rate is positive \cite{ssll}. This is a key step which ensures
relaxation to local equilibrium, and leads to important constraints on the transport co-efficients. In $d=2$, it was found that situations with the velocity $u^\mu$ spacetime independent are characterized
by only a {\em single\/} independent transport co-efficient \cite{ssnernst}. This we choose to be the longitudinal conductivity
at $B=0$.  
\item Solve the initial value problem for the state variables using the linearized equations
of motion. 
\item Finally, translate this solution to the linear response functions, as described in Ref.~\cite{sskm}.
\end{enumerate}

\subsection{Dyonic black hole}
\label{sec:ssdyon}

Given the success of the AdS/CFT correspondence for the specific supersymmetric model in Section~\ref{sec:ssexact},
we boldly assume a similar correspondence for a generic CFT3. We assume that each CFT3 is dual to a strongly-coupled
theory of gravity on AdS$_4$. Furthermore, given the operators associated with the perturbations away from the pure
CFT3 we want to study, we can also deduce the corresponding perturbations away from the dual gravity theory.
So far, this correspondence is purely formal and not of much practical use to us. However, we now restrict our
attention to the hydrodynamic, collision dominated regime, $\hbar \omega \ll k_B T$ of the CFT3. We would like
to know the corresponding low energy effective theory describing the quantum gravity theory on AdS$_4$. 
Here, we make the simplest possible assumption: the effective theory is just the Einstein-Maxwell theory
of general relativity and electromagnetism on AdS$_4$. As in Section~\ref{sec:ssexact}, the temperature $T$ of CFT3
corresponds to introducing a black hole on AdS$_4$ whose Hawking temperature is $T$. The chemical potential, $\mu$,
of the CFT3 corresponds to an electric charge on the black hole, and the applied magnetic field maps to a 
magnetic charge on the black hole. Such a dynoic black hole solution of the Einstein-Maxwell equations is, in fact, known:
it is the Reissner-Nordstrom black hole.

We solved the classical Einstein-Maxwell equations for linearized fluctuations about the metric of a dyonic black hole
in a space which is asymptotically AdS$_4$. The results were used to obtain correlators of a CFT3 using 
the prescriptions of the AdS/CFT mapping.
As we have noted,
we have no detailed knowledge of the strongly-coupled quantum gravity theory which is dual to the CFT3 describing
the superfluid-insulator transition in condensed matter systems, or of graphene. 
Nevertheless, given our postulate that its low
energy effective field theory essentially captured by the Einstein-Maxwell theory, we can then obtain a powerful set
of results for CFT3s. 

\subsection{Results}
\label{sec:ssresults}

In the end, we obtained complete agreement between the two independent computations in Sections~\ref{sec:ssmagneto}
and~\ref{sec:ssdyon}, after allowing for their distinct equations of state. This agreement demonstrates that
the assumption of a low energy Einstein-Maxwell effective field theory for a strongly coupled theory of quantum gravity
is equivalent to the assumption of hydrodynamic transport for $\hbar \omega \ll k_B T$ in a strongly coupled CFT3.

Finally, we turn to our explicit results for quantum critical transport with $\hbar \omega \ll k_B T$. 

First, consider adding a chemical potential, $\mu$,  to the CFT3. This will induce a non-zero
number density of carriers $\rho$. The value of $\rho$ is defined so that the total charge density associated with
$\rho$ is $e^{\ast} \rho$. Then the electrical conductivity at a frequency $\omega$ is 
\begin{equation}
\sigma (\omega) = \frac{e^{\ast 2}}{h} \Phi_\sigma + \frac{e^{\ast 2} \rho^2 v^2}{(\varepsilon + P)} \frac{1}{(- i \omega + 1/\tau_{\rm imp})}
\label{sssw}
\end{equation}
In this section, we are again using the symbol $v$ to denote the characteristic velocity of the CFT3 because we will need $c$
for the physical velocity of light below. Here $\varepsilon$ is the energy density and $P$ is the pressure of the CFT3.
We have assumed a small density of impurities which lead to a momentum relaxation time $\tau_{\rm imp}$ \cite{ssnernst,sshh}.
In general, $\Phi_\sigma$, $\rho$, $\varepsilon$, $P$, and $1/\tau_{\rm imp}$ will be functions of $\mu/k_B T$
which cannot be computed by hydrodynamic considerations alone. However, apart from $\Phi_\sigma$, these quantities
are usually amenable to direct perturbative computations in the CFT3, or by quantum Monte Carlo studies.
The physical interpretation of Eq.~(\ref{sssw}) should be evident: adding a charge density $\rho$ leads to an 
additional Drude-like contribution to the conductivity. This extra current cannot be relaxed by collisions between the 
unequal density of particle and hole excitations, and so requires an impurity relaxation mechanism to yield a finite
conductivity in the d.c. limit.

Now consider thermal transport in a CFT3 with a non-zero $\mu$. The d.c. thermal conductivity, $\kappa$, is given by
\begin{equation}
\kappa = \Phi_\sigma \left( \frac{k_B^2 T}{h} \right) \left( \frac{\varepsilon + P}{k_B T \rho}
\right)^2 , \label{sskapparho}
\end{equation}
in the absence of impurity scattering, $1/\tau_{\rm imp} \rightarrow 0$. This is a Wiedemann-Franz-like relation, connecting the thermal conductivity to the electrical conductivity
in the $\mu=0$ CFT. Note that $\kappa$ diverges as $\rho \rightarrow 0$, 
and so the thermal conductivity of the $\mu=0$ CFT is infinite.

Next, turn on a small magnetic field $B$; we assume that $B$ is small enough
that the spacing between the Landau levels is not as large as $k_B T$. The case
of large Landau level spacing is also experimentally important, but cannot be addressed
by the present analysis. Initially, consider the case $\mu=0$. In this case, the result
Eq.~(\ref{sskapparho}) for the thermal conductivity is replaced by
\begin{equation}
\kappa = \frac{1}{\Phi_\sigma} 
\left( \frac{k_B^2 T}{h} \right) \left( \frac{\varepsilon + P}{k_B T B/(hc/e^\ast)}
\right)^2 \label{sskappaB}
\end{equation}
also in the absence of impurity scattering, $1/\tau_{\rm imp} \rightarrow 0$.
This result relates $\kappa$ to the electrical {\em resistance} at criticality,
and so can be viewed as Wiedemann-Franz-like relation for the vortices.
A similar $1/B^2$ dependence of $\kappa$ appeared in the Boltzmann equation
analysis of Ref.~\cite{ssbgs}, but our more general analysis applies in a wider and distinct regime
\cite{ssmarkus}, and relates the
co-efficient to other observables.

We have obtained a full set of results for the frequency-dependent thermo-electric
response functions at non-zero $B$ and $\mu$. The results are lengthy and we refer
the reader to Ref.~\cite{ssnernst} for explicit expressions. Here we only note that the characteristic feature  \cite{ssnernst,sssean2} of these results is a new {\em hydrodynamic cyclotron resonance}.
The usual cyclotron resonance occurs at the classical cyclotron frequency, which is independent of the particle density and temperature; 
further, in a Galilean-invariant system this resonance
is not broadened by electron-electron interactions alone, and requires impurities
for non-zero damping. The situtation for our hydrodynamic resonance is very different.
It occurs in a collision-dominated regime, and its frequency depends on 
the density and temperature: the explicit expression for the resonance frequency is
\begin{equation}
\omega_c = \frac{e^\ast B \rho v^2}{c (\varepsilon + P)}.
\end{equation}
Further, the cyclotron resonance involves particle and hole excitations moving
in opposite directions, and collisions between them can damp the resonance
even in the absence of impurities. Our expression for this intrinsic damping frequency is \cite{ssnernst,sssean2}
\begin{equation}
\gamma = \frac{e^{\ast 2}}{h} \Phi_\sigma \frac{ B^2 v^2}{c^2 (\varepsilon + P)},
\end{equation}
relating it to the quantum-critical conductivity as a measure of collisions between 
counter-propagating particles and holes.
We refer the reader to a separate discussion \cite{ssmarkus} of the experimental conditions under
which this hydrodynamic cyclotron resonance may be observed.

\section{$d$-wave superconductors}
\label{sec:ssdwave}

We now turn to the second class of strong-coupling problems outlined in Section~\ref{sec:ssintro}:
those involving quantum critical points with fermionic excitations. 
This section will consider the simpler class of problems in which the fermions have
a Dirac spectrum, and the field-theoretic $1/N$ expansion does allow for 
substantial progress.

We will begin in Section~\ref{sec:ssdirac} by an elementary discussion of the origin of these
Dirac fermions. Then we will consider two quantum phase transitions, both involving a simple 
Ising order parameter. The first in Section~\ref{sec:ssdid}, with time-reversal symmetry breaking, leads to a relativistic 
quantum field theory closely related to the Gross-Neveu model. The second model of Section~\ref{sec:ssisingnematic}
involves breaking of a lattice rotation symmetry, leading to ``Ising-nematic'' order. The theory for this model
is not relativistically invariant: it is strongly coupled, but can be controlled by a traditional $1/N$ expansion.

We note that symmetry breaking transitions in graphene are also described by field theories
similar to those discussed in this section \cite{ssherbut,ssherbut2}.

\subsection{Dirac fermions}
\label{sec:ssdirac}
We begin with a review of the standard BCS mean-field
theory for a $d$-wave superconductor on the square lattice, with an eye towards identifying
the fermionic Bogoliubov quasiparticle excitations. For now, we assume we are far from
any QPT associated with SDW, Ising-nematic, or other broken symmetries.
We
consider the Hamiltonian
\begin{equation}
H_{tJ} = \sum_{k} \varepsilon_k c_{k \alpha}^{\dagger} c_{k
\alpha} + J_1 \sum_{\langle ij \rangle} {\bf S}_i \cdot {\bf
S}_{j} \label{g2}
\end{equation}
where $c_{j\alpha}$ is the annihilation operator for an electron
on site $j$ with spin $\alpha=\uparrow,\downarrow$, $c_{k\alpha}$
is its Fourier transform to momentum space, $\varepsilon_k$ is the
dispersion of the electrons (it is conventional to choose
$\varepsilon_k = -2t_1 (\cos(k_x) + \cos(k_y)) - 2 t_2 ( \cos(k_x
+ k_y) + \cos(k_x - k_y)) - \mu$, with $t_{1,2}$ the first/second
neighbor hopping and $\mu$ the chemical potential), and the $J_1$
term is similar to that in Eq.~(\ref{eq:ssHJ}) with
\begin{equation}
S_{ja} = \frac{1}{2} c^{\dagger}_{j\alpha}
\sigma_{\alpha\beta}^{a} c_{j \beta} \label{g2a}
\end{equation}
and $\sigma^{a}$ the Pauli matrices. We will consider
the consequences of the further neighbor exchange interactions for the
superconductor in Section~\ref{sec:ssdid} below. Applying the BCS
mean-field decoupling to $H_{tJ}$ we obtain the Bogoliubov
Hamiltonian
\begin{equation}
H_{BCS} = \sum_{k} \varepsilon_k c_{k \alpha}^{\dagger} c_{k
\alpha} - \frac{J_1}{2} \sum_{j\mu}\Delta_{\mu} \left(
c^{\dagger}_{j\uparrow} c^{\dagger}_{j+\hat{\mu},\downarrow} -
c^{\dagger}_{j\downarrow} c^{\dagger}_{j+\hat{\mu},\uparrow}
\right) + \mbox{h.c.}. \label{g3}
\end{equation}
For a wide range of parameters, the ground state energy optimized
by a $d_{x^2-y^2}$ wavefunction for the Cooper pairs: this
corresponds to the choice $\Delta_x = - \Delta_y =
\Delta_{x^2-y^2}$. The value of $\Delta_{x^2-y^2}$ is determined
by minimizing the energy of the BCS state
\begin{equation}
E_{BCS} = J_1 |\Delta_{x^2-y^2}|^2 - \int \frac{d^2 k}{4 \pi^2}
\left[ E_k - \varepsilon_k \right] \label{g4}
\end{equation}
where the fermionic quasiparticle dispersion is
\begin{equation}
E_k = \left[ \varepsilon_k^2 + \left|J_1 \Delta_{x^2-y^2}(\cos k_x
- \cos k_y)\right|^2 \right]^{1/2}. \label{g5}
\end{equation}

The energy of the quasiparticles, $E_k$, vanishes at the four
points $(\pm Q, \pm Q)$ at which $\varepsilon_k=0$. We are
especially interested in the low energy quasiparticles in the
vicinity of these points, and so we perform a gradient expansion
of $H_{BCS}$ near each of them. We label the points $\vec{Q}_1=(Q,Q)$,
$\vec{Q}_2=(-Q,Q)$, $\vec{Q}_3=(-Q,-Q)$, $\vec{Q}_4=(Q,-Q)$ and write
\begin{equation}
c_{j\alpha} =f_{1\alpha} (\vec{r}_j) e^{i \vec{Q}_1 \cdot \vec{r}_j}+f_{2\alpha} (\vec{r}_j)
e^{i \vec{Q}_2 \cdot \vec{r}_j}+f_{3\alpha} (\vec{r}_j) e^{i \vec{Q}_3 \cdot \vec{r}_j}+f_{4\alpha} (\vec{r}_j)
e^{i \vec{Q}_4 \cdot \vec{r}_j},\label{g5a}
\end{equation}
while assuming the $f_{1-4,\alpha} (\vec{r})$ are slowly varying
functions of $x$. We also introduce the bispinors $\Psi_1 =
(f_{1\uparrow}, f_{3\downarrow}^{\dagger},
f_{1\downarrow},-f_{3\uparrow}^{\dagger})$, and $\Psi_2 =
(f_{2\uparrow}, f_{4\downarrow}^{\dagger},
f_{2\downarrow},-f_{4\uparrow}^{\dagger})$, and then express
$H_{BCS}$ in terms of $\Psi_{1,2}$ while performing a spatial
gradient expansion. This yields the following effective action for
the fermionic quasiparticles:
\begin{eqnarray}
&& \mathcal{S}_{\Psi} = \int d \tau d^2 r \Biggl[
\Psi_{1}^{\dagger}  \left(
\partial_\tau -i \frac{v_F}{\sqrt{2}} (\partial_x + \partial_y) \tau^z  -i \frac{v_\Delta}{\sqrt{2}} (-\partial_x + 
\partial_y) \tau^x \right) \Psi_{1}   \nonumber \\
&&~~~~~~~~~~~~~~~~~~+  \Psi^\dagger_2 \left(
\partial_\tau - i \frac{v_F}{\sqrt{2}} (-\partial_x + \partial_y) \tau^z -i  \frac{v_\Delta}{\sqrt{2}} (\partial_x + \partial_y)  \tau^x \right) \Psi_{2} \Biggr] .\label{dsid1}
\end{eqnarray}
where the $\tau^{x,z}$ are $4 \times 4$ matrices which
are block diagonal, the blocks consisting of $2\times 2$ Pauli
matrices. The velocities $v_{F,\Delta}$ are given by the conical
structure of $E_k$ near the $Q_{1-4}$: we have $v_F =
\left|\nabla_k \varepsilon_k |_{k=Q_a} \right|$ and $v_{\Delta} =
|J_1 \Delta_{x^2-y^2} \sqrt{2} \sin (Q)|$. In this limit, the energy of the 
$\Psi_1$ fermionic excitations is
$E_k = ( v_F^2 (k_x+k_y)^2 /2 + v_\Delta^2 (k_x - k_y)^2 /2)^{1/2}$ (and similarly for $\Psi_2$), which is the spectrum of massless
Dirac fermions.

\subsection{Time-reversal symmetry breaking}
\label{sec:ssdid}

We will consider a simple model in which the pairing symmetry of the
superconductor changes from $d_{x^2-y^2}$ to $d_{x^2-y^2} \pm i d_{xy}$.
The choice of the phase between the two pairing components
leads to a breaking of time-reversal symmetry. Studies of this transition
were originally motivated by the cuprate phenomenology, but we will not
explore this experimental connection here because the evidence has
remained sparse. 

The mean field theory of this transition can be explored entirely within
the context of BCS theory, as we will review below. However, fluctuations
about the BCS theory are strong, and lead to non-trivial critical behavior
involving both the collective order parameter and the Bogoliubov fermions: 
this is probably the earliest
known example \cite{vojta1,vojta2} of the failure of BCS theory in two (or higher)
dimensions in a superconducting ground state. At $T>0$, this
failure broadens into the ``quantum critical'' region.

We extend $H_{tJ}$ in Eq.~(\ref{g2}) so
that BCS mean-field theory permits a region with $d_{xy}$
superconductivity. With a $J_2$ second neighbor interaction,
Eq.~(\ref{g2}) is modified to:
\begin{equation}
\widetilde{H}_{tJ} = \sum_{k} \varepsilon_k c_{k \sigma}^{\dagger}
c_{k \sigma} + J_1 \sum_{\langle ij \rangle} {\bf S}_i \cdot {\bf
S}_{j} + J_2 \sum_{{\rm nnn}~ij} {\bf S}_i \cdot {\bf S}_j.
\label{g8}
\end{equation}
We will follow the evolution of the ground state of
$\widetilde{H}_{tJ}$ as a function of $J_2 / J_1$. 

\begin{figure}
\centerline{\includegraphics[width=2in]{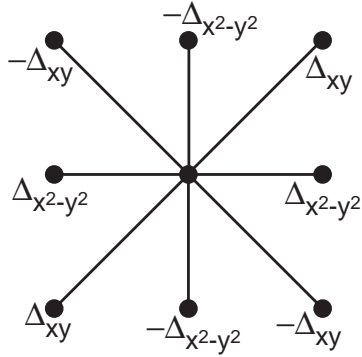}}
\caption{Values of the pairing amplitudes, $-\langle c_{i
\uparrow} c_{j \downarrow} -  c_{i \downarrow} c_{j \uparrow}
\rangle$ with $i$ the central site, and $j$ is one of its 8 near
neighbors.} \label{fig11}
\end{figure}

The
mean-field Hamiltonian is now modified from Eq.~(\ref{g3}) to
\begin{eqnarray}
\widetilde{H}_{BCS} = \sum_{k} \varepsilon_k c_{k
\sigma}^{\dagger} c_{k \sigma} &-& \frac{J_1}{2} \sum_{j,\mu}
\Delta_{\mu} (c_{j\uparrow}^{\dagger}
c_{j+\hat{\mu},\downarrow}^{\dagger} - c_{j\downarrow}^{\dagger}
c_{j+\hat{\mu},\uparrow}^{\dagger}) + \mbox{h.c.} \nonumber \\ &-&
\frac{J_2}{2} {\sum_{j,\nu}}^{\prime} \Delta_{\nu}
(c_{j\uparrow}^{\dagger} c_{j+\hat{\nu},\downarrow}^{\dagger} -
c_{j\downarrow}^{\dagger} c_{j+\hat{\nu},\uparrow}^{\dagger}) +
\mbox{h.c.}, \label{g9}
\end{eqnarray}
where the second summation over $\nu$ is along the diagonal
neighbors $\hat{x}+\hat{y}$ and $-\hat{x}+\hat{y}$. To obtain
$d_{xy}$ pairing along the diagonals, we choose $\Delta_{x+y} = -
\Delta_{-x+y} = \Delta_{xy}$. We summarize our choices for the
spatial structure of the pairing amplitudes (which determine the
Cooper pair wavefunction) in Fig~\ref{fig11}.
The values of $\Delta_{x^2-y^2}$ and $\Delta_{xy}$ are to be
determined by minimizing the ground state energy (generalizing
Eq.~(\ref{g4}))
\begin{equation}
E_{BCS} = J_1 |\Delta_{x^2-y^2}|^2 +J_2 |\Delta_{xy}|^2 - \int
\frac{d^2 k}{4 \pi^2} \left[ E_k - \varepsilon_k \right]
\label{g10}
\end{equation}
where the quasiparticle dispersion is now (generalizing
Eq.~(\ref{g5}))
\begin{equation}
E_k = \left[ \varepsilon_k^2 + \left|J_1 \Delta_{x^2-y^2}(\cos k_x
- \cos k_y) + 2 J_2 \Delta_{xy} \sin k_x \sin k_y \right|^2
\right]^{1/2}. \label{g11}
\end{equation}
Notice that the energy depends upon the relative phase of
$\Delta_{x^2-y^2}$ and $\Delta_{xy}$: this phase is therefore an
observable property of the ground state.

\begin{figure}[t]
\centerline{\includegraphics[width=4in]{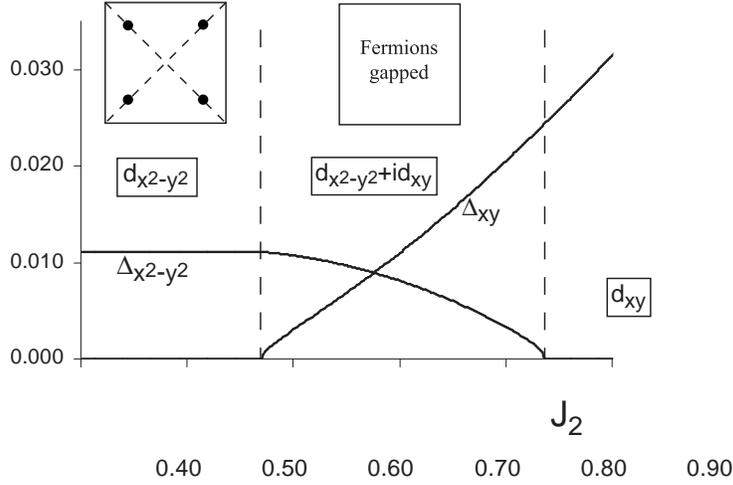}} \caption{BCS
solution of the phenomenological Hamiltonian $\widetilde{H}_{tJ}$
in Eq.~(\ref{g8}). Shown are the optimum values of the pairing
amplitudes $|\Delta_{x^2-y^2}|$ and $|\Delta_{xy}|$ as a function
of $J_2$ for $t_1 =1$, $t_2=-0.25$, $\mu=-1.25$, and $J_1$ fixed
at $J_1=0.4$. The relative phase of the pairing amplitudes was
always found to obey Eq.~(\ref{g12}). The dashed lines denote
locations of phase transitions between $d_{x^2-y^2}$,
$d_{x^2-y^2}+id_{xy}$, and $d_{xy}$ superconductors. The pairing
amplitudes vanishes linearly at the first transition
corresponding to the exponent $\beta_{BCS}=1$ in
Eq.~(\ref{g14}). The Brillouin zone location of the gapless Dirac points 
in the $d_{x^2-y^2}$ superconductor is indicated by filled circles. 
For the dispersion $\varepsilon_k$ appropriate to
the cuprates, the $d_{xy}$ superconductor is fully gapped, and so the 
second transition is ordinary Ising. } \label{fig12}
\end{figure}

It is a simple matter to numerically carry out the minimization of
Eq.~(\ref{g11}), and the results for a typical choice of parameters
are shown in Fig~\ref{fig12} as a function $J_2/J_1$.
One of the two amplitudes $\Delta_{x^2-y^2}$ or $\Delta_{xy}$ is
always non-zero and so the ground state is always superconducting.
The transition from pure $d_{x^2-y^2}$ superconductivity to pure
$d_{xy}$ superconductivity occurs via an intermediate phase in
which {\em both} order parameters are non-zero. Furthermore, in
this regime, their relative phase is found to be pinned to $\pm
\pi/2$ {\em i.e.}
\begin{equation}
\arg (\Delta_{xy}) = \arg (\Delta_{x^2-y^2}) \pm \pi/2 \label{g12}
\end{equation}
The reason for this pinning can be intuitively seen from
Eq.~(\ref{g11}): only for these values of the relative phase does the
equation $E_k = 0$ never have a solution. In other words, the
gapless nodal quasiparticles of the $d_{x^2-y^2}$ superconductor
acquire a finite
energy gap when a secondary pairing with relative phase $\pm
\pi/2$ develops. By a level repulsion picture, we can expect that
gapping out the low energy excitations should help lower the
energy of the ground state. The intermediate phase obeying
Eq.~(\ref{g12}) is called a $d_{x^2-y^2} + i d_{xy}$ superconductor.

The choice of the sign in Eq.~(\ref{g12}) leads to an overall two-fold
degeneracy in the choice of the wavefunction for the $d_{x^2-y^2}
+ i d_{xy}$ superconductor. This choice is related to the breaking
of time-reversal symmetry, and implies that the $d_{x^2-y^2} + i
d_{xy}$ phase is characterized by the non-zero expectation value
of a $Z_2$ Ising order parameter; the expectation value of this
order vanishes in the two phases (the $d_{x^2-y^2}$ and $d_{xy}$
superconductors) on either side of the $d_{x^2-y^2}+id_{xy}$
superconductor. As is conventional, we will represent the Ising
order by a real scalar field $\phi$. Fluctuations of $\phi$ become
critical near both of the phase boundaries in Fig~\ref{fig12}. As
we will explain below, the critical theory of the $d_{x^2-y^2}$ to
$d_{x^2-y^2} + i d_{xy}$ transition is {\em not} the usual $\phi^4$ field
theory which describes the ordinary Ising transition in three
spacetime dimensions. (For the dispersion $\varepsilon_k$ appropriate to
the cuprates, the $d_{xy}$ superconductor is fully gapped, and so the 
$d_{x^2-y^2} + i d_{xy}$ to $d_{xy}$ transition in Fig.~\ref{fig12} will be 
ordinary Ising.)

Near the phase boundary from $d_{x^2-y^2}$ to $d_{x^2-y^2} +
id_{xy}$ superconductivity it is clear that we can identify
\begin{equation}
\phi = i \Delta_{xy}, \label{o1}
\end{equation}
(in the gauge where $\Delta_{x^2-y^2}$ is real). We can now expand
$E_{BCS}$ in Eq.~(\ref{g10}) for small $\phi$ (with $\Delta_{x^2-y^2}$
finite) and find a series with the structure~\cite{laugh,wolf}
\begin{equation}
E_{BCS} = E_0 + s \phi^2 + v |\phi|^3 + \ldots, \label{g13}
\end{equation}
where $s$, $v$ are coefficients and the ellipses represent regular
higher order terms in even powers of $\phi$; $s$ can have either
sign, whereas $v$ is always positive. Notice the non-analytic
$|\phi|^3$ term that appears in the BCS theory
--- this arises from an infrared singularity in the integral in
Eq.~(\ref{g10}) over $E_k$ at the four nodal points of the
$d_{x^2-y^2}$ superconductor, and is a preliminary indication that
the transition differs from that in the ordinary Ising model, and that the Dirac
fermions play a central role. We can optimize
$\phi$ by minimizing $E_{BCS}$ in Eq.~(\ref{g13})--- this shows that
$\langle \phi \rangle =0$ for $s>0$, and $\langle \phi \rangle
\neq 0$ for $s<0$. So $s \sim (J_2/J_1)_c - J_2/J_1$ where
$(J_2/J_1)_c$ is the first critical value in Fig~\ref{fig12}. Near
this critical point, we find
\begin{equation}
\langle \phi \rangle \sim (s_c - s)^{\beta}, \label{g14}
\end{equation}
where we have allowed for the fact that fluctuation corrections
will shift the critical point from $s=0$ to $s=s_c$. The present
BCS theory yields the exponent $\beta_{BCS} = 1$; this differs
from the usual mean-field exponent $\beta_{MF} = 1/2$, and this is
of course due to the non-analytic $|\phi|^3$ term in Eq.~(\ref{g13}).

We can now write down the required field theory
of the onset of $d_{xy}$ order. In addition to the order parameter $\phi$,
the field theory should also involve the low energy nodal fermions
of the $d_{x^2-y^2}$ superconductor, as described by
$\mathcal{S}_{\Psi}$ in Eq.~(\ref{dsid1}). For the $\phi$ fluctuations,
we write down the usual terms permitted near a phase transition
with Ising symmetry:
\begin{equation}
\mathcal{S}_{\phi} = \int d^2 r d\tau \left[\frac{1}{2} \left(
(\partial_{\tau} \phi)^2 + c^2 (\partial_x \phi)^2 + c^2
(\partial_y \phi)^2 + s \phi^2 \right) + \frac{u}{24} \phi^4
\right]. \label{g15}
\end{equation}
Note that, unlike Eq.~(\ref{g13}), we do not have any non-analytic
$|\phi|^3$ terms in the action: this is because we have not
integrated out the low energy Dirac fermions, and the terms in
Eq.~(\ref{g15}) are viewed as arising from high energy fermions away
from the nodal points. Finally, we need to couple the $\phi$ and
$\Psi_{1,2}$ excitations. Their coupling is already contained in
the last term in Eq.~(\ref{g9}): expressing this in terms of the
$\Psi_{1,2}$ fermions using Eq.~(\ref{g5a}) we obtain
\begin{equation}
\mathcal{S}_{\Psi \phi} = \vartheta_{xy} \int d^2 r d \tau \left[  \phi
\left( \Psi_1^{\dagger} \tau^y \Psi_1 - \Psi_2^{\dagger} \tau^y
\Psi_2 \right) \right], \label{g16}
\end{equation}
where $\vartheta_{xy}$ is a coupling constant. The partition function of
the full theory is now
\begin{equation}
\mathcal{Z}_{did} = \int \mathcal{D}\phi \mathcal{D} \Psi_1 \mathcal{D}
\Psi_2 \exp \left( - \mathcal{S}_{\Psi} - \mathcal{S}_{\phi} -
\mathcal{S}_{\Psi \phi} \right), \label{g17}
\end{equation}
where $\mathcal{S}_{\Psi}$ was in Eq.~(\ref{dsid1}). It can now be
checked that if we integrate out the $\Psi_{1,2}$ fermions for a
spacetime independent $\phi$, we do indeed obtain a $|\phi|^3$
term in the effective potential for $\phi$.

We begin our analysis of $Z_{did}$ by assuming that the transition
is described by a fixed point with $\vartheta_{xy}=0$: then 
the theory for the transition would be the
ordinary $\phi^4$ field theory $\mathcal{S}_{\phi}$, and the nodal
fermions would be innocent spectators.
The scaling
dimension of $\phi$ at such a fixed point is $(1 + \eta_I )/2$
(where $\eta_I$ is the anomalous order parameter exponent at the
critical point of the ordinary three dimensional Ising model),
while that of $\Psi_{1,2}$ is 1. Consequently, the scaling
dimension of $\vartheta_{xy}$ is $(1-\eta_I)/2 > 0$. This positive scaling dimension
implies that $\vartheta_{xy}$ is relevant and the $\vartheta_{xy}=0$ fixed point
is unstable: the Dirac fermions are fully involved in the critical theory.

Determining the correct critical behavior now requires a full
renormalization group analysis of $Z_{did}$. This has been
described in some detail in Ref.~\cite{vojta2}, and we will not
reproduce the details here. The main result we need for our
purposes is that couplings $\vartheta_{xy}$, $u$, $v_F /c$ and
$v_{\Delta}/c$ all reach {\em non-zero} fixed point values which
define a critical point in a new universality class. These fixed
point values, and the corresponding critical exponents, can be
determined in expansions in either $(3-d)$~\cite{vojta1,vojta2}
(where $d$ is the spatial dimensionality) or $1/N$~\cite{kvesch}
(where $N$ is the number of fermion species). 
An important simplifying feature here is that the fixed point is actually
relativistically invariant.
Indeed the fixed
point has the structure of the so-called Higgs-Yukawa (or Gross-Neveu) model which
has been studied extensively in the particle physics
literature~\cite{baruch} in a different physical context: quantum
Monte Carlo simulation of this model also exist~\cite{qmc}, and
provide probably the most accurate estimate of the exponents.

The non-trivial fixed point has strong implications for
the correlations of the Bogoliubov fermions. The
fermion correlation function $G_1 = \langle \Psi_1
\Psi_1^{\dagger} \rangle$ obeys
\begin{equation}
G_1 (k, \omega) = \frac{\omega + v_F k_x \tau^z + v_{\Delta}
\tau^x}{(v_F^2 k_x^2 + v_{\Delta}^2 k_y^2 -
\omega^2)^{(1-\eta_f)/2}} \label{g18}
\end{equation}
at low frequencies for $s \geq s_c$. Away from the critical point
in the $d_{x^2-y^2}$ superconductor with $s>s_c$, Eq~(\ref{g18})
holds with $\eta_f = 0$, and this is the BCS result, with sharp
quasi-particle poles in the Green's function. At the critical
point $s=s_c$ Eq.~(\ref{g18}) holds with the fixed point values for
the velocities (which satisfy $v_F = v_{\Delta} = c$) and with the
anomalous dimension $\eta_f \neq 0$
--- the $(3-d)$ expansion~\cite{vojta1} estimate is $\eta_f
\approx (3-d)/14$, and the $1/N$ expansion estimate~\cite{kvesch}
is $\eta_f \approx 1/(3 \pi^2 N)$, with $N=2$. This is clearly
non-BCS behavior, and the fermionic quasiparticle pole in the
spectral function has been replaced by a branch-cut representing
the continuum of critical excitations. The corrections to BCS
extend also to correlations of the Ising order $\phi$: its
expectation value vanishes as Eq.~(\ref{g14}) with the Monte Carlo
estimate $\beta \approx 0.877$~\cite{qmc}. The critical point
correlators of $\phi$ have the anomalous dimension $\eta
\approx 0.754$~\cite{qmc}, which is clearly different from the
very small value of the exponent $\eta_I$ at the unstable
$\vartheta_{xy}=0$ fixed point. The value of $\beta$ is related to $\eta$
by the usual scaling law $\beta = (1+\eta) \nu/2$, with $\nu
\approx 1.00$ the correlation length exponent (which also differs
from the exponent $\nu_I$ of the Ising model).

\subsection{Nematic ordering}
\label{sec:ssisingnematic}

We now consider an Ising transition associated with ``Ising-nematic'' ordering in the $d$-wave
superconductor. This is associated with a spontaneous reduction of the lattice symmetry of the 
Hamiltonian from ``square'' to ``rectangular''. Our study is motivated by experimental observations
of such a symmetry breaking in the cuprate superconductors \cite{ando02,hinkov08a,louisnematic}.

The ingredients of such an ordering are actually already present in our simple review of 
BCS theory in Section~\ref{sec:ssdirac}. In Eq.~(\ref{g3}), we introduce 2 variational pairing amplitudes
$\Delta_x$ and $\Delta_y$. Subsequently, we assumed that the minimization of the energy
led to a solution with $d_{x^2-y^2}$ pairing symmetry with $\Delta_x = - \Delta_y = \Delta_{x^2 - y^2}$.
However, it is possible that upon including the full details of the microscopic interactions we
are led to a minimum where the optimal solution also has a small amount of $s$-wave pairing.
Then $|\Delta_x| \neq |\Delta_y|$, and we can expect all physical properties to have distinct
dependencies on the $x$ and $y$ co-ordinates. 
Thus, one measure of the the Ising nematic order parameter is $|\Delta_x|^2 - |\Delta_y|^2$.

\begin{figure}[t]
\centerline{\includegraphics[width=3.5in]{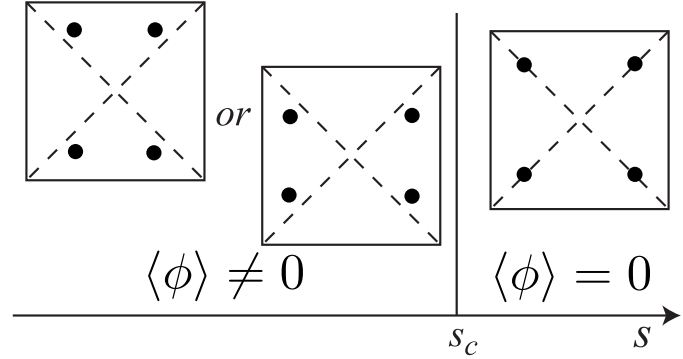}} \caption{
Phase diagram of Ising nematic ordering in a $d$-wave superconductor as a function 
of the coupling $s$ in $\mathcal{S}_\phi$. The filled circles indicate the location
of the gapless fermionic excitations in the Brillouin zone. The two choices for $s<s_c$
are selected by the sign of $\langle \phi \rangle.$} \label{fig:dnematic}
\end{figure}

The derivation of the field theory for this transition follows closely our presentation in 
Section~\ref{sec:ssdid}. We allow for small Ising-nematic ordering by introducing a scalar field $\phi$
and writing
\begin{equation}
\Delta_x = \Delta_{x^2 - y^2} + \phi~~~;~~~\Delta_y = - \Delta_{x^2-y^2} + \phi .
\end{equation}
The evolutions of the Dirac fermion spectrum under such a change is indicated
in Fig.~\ref{fig:dnematic}.
We now develop an effective action for $\phi$ and the Dirac fermions $\Psi_{1,2}$.
The result is essentially identical to that in Section~\ref{sec:ssdid}, apart from a change
in the structure of the Yukawa coupling. Thus we obtain a theory
$\mathcal{S}_\Psi + \mathcal{S}_\phi + \overline{\mathcal{S}}_{\Psi\phi}$, 
defined by Eqs.~(\ref{dsid1}) and (\ref{g15}), and
where
Eq.~(\ref{g16}) is now replaced by
\begin{equation}
\overline{\mathcal{S}}_{\Psi \phi} = \vartheta_{I} \int d^2 r d \tau \left[  \phi
\left( \Psi_1^{\dagger} \tau^x \Psi_1 + \Psi_2^{\dagger} \tau^x
\Psi_2 \right) \right]. \label{h16}
\end{equation}

The seemingly innocuous change between Eqs.~(\ref{g16}) and (\ref{h16}) however
has strong consequences. This is partly linked to the fact with $\overline{\mathcal{S}}_{\Psi\phi}$
cannot be relativistically invariant even after all velocities are adjusted to equal.
A weak-coupling renormalization group analysis in powers of the coupling $\vartheta_I$ was performed
in $(3-d)$ dimensions in Refs.~\cite{vojta1,vojta2}, and led to flows to strong coupling with no
accessible fixed point: thus no firm conclusions on the nature of the critical theory
were drawn.

This problem remained unsolved until the recent works of Refs.~\cite{kim,yejin}.
It is essential that there not be any expansion in powers of the coupling $\vartheta_I$.
This is because it leads to strongly non-analytic changes in the structure of the $\phi$
propagator, which have to be included at all stages. In a model with $N$ fermion flavors,
the $1/N$ expansion does avoid any expansion in $\vartheta_I$. The renormalization group
analysis has to be carried out within the context of the $1/N$ expansion, and this
involves some rather technical analysis which is explained in Ref.~\cite{yejin}. 
In the end, an asymptotically exact description of the vicinity of the critical point was obtained.
It was found that the velocity ratio $v_F / v_\Delta$ diverged logarithmically with energy scale,
leading to strongly anisotropic `arc-like' spectra for the Dirac fermions. Associated singularities
in the thermal conductivity have also been computed \cite{lars}.

\section{Metals}
\label{sec:ssmetal}

This section considers symmetry breaking transitions in two-dimensional metals.
Away from the quantum critical point, the phases will be ordinary Fermi liquids.
We will be interested in the manner in which the Fermi liquid behavior breaks down
at the quantum critical point. Our focus will be exclusively on two spatial dimensions:
quantum phase transitions of metals in three dimensions are usually simpler,
and the traditional perturbative theory appears under control.

In Section~\ref{sec:ssdwave} the fermionic excitations had vanishing energy only 
at isolated nodal points in the Brillouin zone: see Figs.~\ref{fig12} and~\ref{fig:dnematic}. Metals
have fermionic excitations with vanishing energy
along a {\em line\/} in the Brillouin zone. Thus we can expect them to have an even 
stronger effect on the critical theory. This will indeed be the case, and we will be led to problems
with a far more complex structure. Unlike the situation in insulators and $d$-wave superconductor,
many basic issues associated with ordering transition in two dimensional metals have not been
fully resolved. The problem remains one of active research and is being addressed by
many different approaches. In  recent papers \cite{max1,max2}, Metlitski and the author have argued
that the problem is strongly coupled, and proposed field theories and scaling structures
for the vicinity of the critical point. We will review the main ingredients for the transition involving
Ising-nematic ordering in a metal. Thus the symmetry breaking will be just as in 
Section~\ref{sec:ssisingnematic}, but the fermionic spectrum will be quite different.
Our study here is also motivated by experimental observations 
in the cuprate superconductors \cite{ando02,hinkov08a,louisnematic}.

\begin{figure}[t]
\centerline{\includegraphics[width=3.5in]{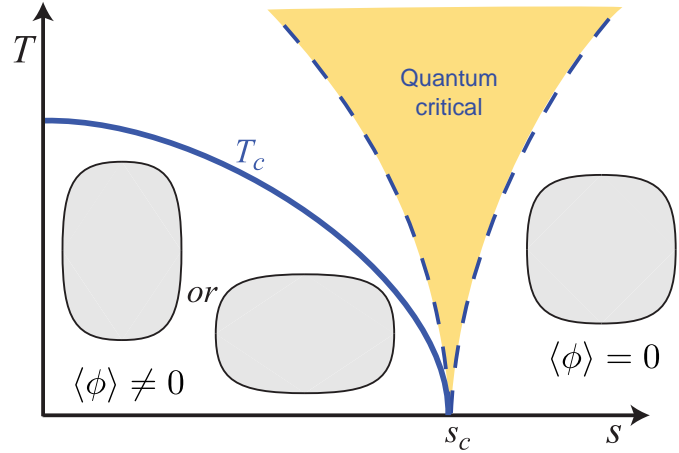}} \caption{
Phase diagram of Ising nematic ordering in a metal as a function 
of the coupling $s$ in $\mathcal{S}_\phi$ and temperature $T$. The Fermi surface for $s>0$ is 
as in the overdoped region of the cuprates, with the shaded region indicating the occupied hole (or empty electron) states. 
The choice between the two quadrapolar distortions
of the Fermi surface is determined by the sign of $\langle \phi \rangle$.
The line of $T>0$ phase transitions at $T_c$ is described by Onsager's solution of the classical
two-dimensional Ising model. We are interested here in the quantum critical point at $s=s_c$,
which controls the quantum-critical region.
} \label{fig:fnematic}
\end{figure}

As in Section~\ref{sec:ssdwave}, let us begin by a description of the non-critical fermionic
sector, before its coupling to the order parameter fluctuations.
We use the band structure describing the cuprates in the over-doped region, well away from the Mott insulator. Here the electrons
$c_{\vec{k} \alpha}$ are described by the kinetic energy in Eq.~(\ref{g2}), which we write in the
following action
\begin{equation}
\mathcal{S}_c = \int d\tau \sum_{\vec{k}} c_{k \alpha}^\dagger \left( \frac{\partial}{\partial \tau} + \varepsilon_k
\right) c_{k \alpha}, \label{sc}
\end{equation}
As in Section~\ref{sec:ssisingnematic}, we will have an 
Ising order parameter represented by the real scalar field $\phi$,
which is described as before by $\mathcal{S}_\phi$ in Eq.~(\ref{g15}). 
Its coupling to the electrons can be deduced by
symmetry considerations, and the most natural coupling (the analog of Eqs.~(\ref{h16})) is
\begin{equation}
\mathcal{S}_{c \phi} = \frac{1}{V} \int d \tau \sum_{\vec{k}, \vec{q}} (\cos k_x - \cos k_y ) \phi (\vec{q}) c_{\vec{k}+\vec{q}/2, \alpha}^\dagger 
c_{\vec{k}-\vec{q}/2, \alpha}.
\label{sci}
\end{equation}
where $V$ is the volume. The momentum dependent form factor is the simplest choice which changes sign under $x \leftrightarrow y$, as is required
by the symmetry properties of $\phi$. The sum over $\vec{q}$ is over small momenta, 
while that over $\vec{k}$
extends over the entire Brillouin zone. The theory for the nematic ordering transition is now
described by $\mathcal{S}_c + \mathcal{S}_\phi +  \mathcal{S}_{c \phi}$. The 
phase diagram as a function of the coupling $s$ in $\mathcal{S}_\phi$ and temperature $T$ 
is shown in Fig.~\ref{fig:fnematic}.
Note that there is a line of Ising phase transitions at $T=T_c$: this transition is in the same universality
class as the classical two-dimensional Ising model. However, quantum effects and fermionic excitations
are crucial at $T=0$ critical point at $s=s_c$ and its associated quantum critical region.

A key property of Eq.~(\ref{sci}) is that small momentum critical $\phi$ fluctuations can efficiently scatter
fermions at every point on the Fermi surface. Thus the non-Fermi singularities in the fermion
Green's function will extend to all points on the Fermi surface. This behavior is dramatically different
from all the field theories we have met so far, all of which had singularities only at isolated points in 
momentum space. We evidently have to write down a long-wavelength theory which has singularities
along a line in momentum space.

\begin{figure}[t]
\centerline{\includegraphics[width=2.5in]{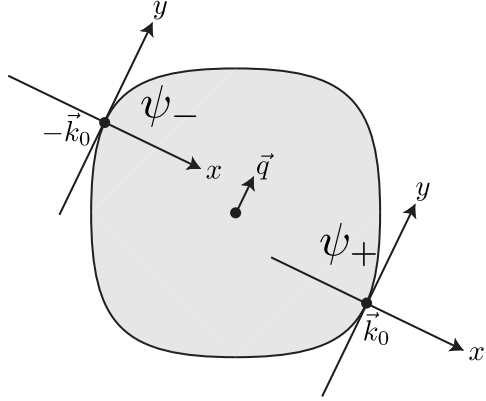}} \caption{
A $\phi$ fluctuation at wavevector $\vec{q}$ couples most efficiently to the 
fermions $\psi_\pm$ near the Fermi surface points $\pm \vec{k}_0$.
} \label{fig:ssfs}
\end{figure}

We describe the construction of Ref.~\cite{max1} of a field theory with this unusual property.
Pick a fluctuation of the order parameter $\phi$ at a momentum $\vec{q}$. As shown in Fig.~\ref{fig:ssfs}
this fluctuation will couple most efficiently to fermions near two points on the Fermi surface,
where the tangent to the Fermi surface is parallel to $\vec{q}$. A fermion absorbing momentum
$\vec{q}$  at these points, changes its energy only by $\sim \vec{q}^2$; at all other points
on the Fermi surface the change is $\sim \vec{q}$.
Thus we are led to focus on different points on the Fermi surface for each direction of $\vec{q}$. 
In the continuum limit, we will therefore need a separate field theory for each pair of points $\pm \vec{k}_0$
on the Fermi surface. We conclude that the quantum critical point is described by an 
infinite number of field theories.

There have been earlier descriptions of Fermi surfaces by an infinite number of field 
theories \cite{Luther,Marston,KwonMarston,haldane,NetoFradkin,LFBFO06,LF07}. However
many of these earlier works differ in a crucial respect from the the theory to be presented here. They focus
on the motion of fermions transverse to Fermi surface, and so represent each Fermi surface
point by a 1+1 dimensional chiral fermion. Thus they have an infinite number of 1+1 dimensional
field theories, labeled by points on the one dimensional Fermi surface. The original problem was
2+1 dimensional, and so this conserves the total dimensionality and the number of degrees of freedom.
However, we have already argued above that the dominant fluctuations of the fermions are
in a direction transverse to the Fermi surface, and so we believe this earlier approach is not
suited for the vicinity of the nematic quantum critical point. As we will see in Section~\ref{sec:ssfield}, 
we will need
an infinite number of 2+1
dimensional field theories labeled by points on the Fermi surface. (Some of the earlier works included fluctuations
transverse to the Fermi surface \cite{LFBFO06,LF07}, but did not account for the curvature of the Fermi surface;
it is important to take the scaling limit at fixed curvature, as we will see.) Thus we have an emergent
dimension and a redundant description of the degrees of freedom. We will see in 
Section~\ref{sec:sssym} how 
compatibility conditions ensure that the redundancy does not lead to any inconsistencies. 
The emergent dimensionality suggests
a connection to the AdS/CFT correspondence, as will be discussed in Section~\ref{sec:ssadsmetal}.

\subsection{Field theories}
\label{sec:ssfield}

Let us now focus on the vicinity of the points $\pm \vec{k}_0$, by introducing fermionic
field $\psi_\pm$ by
\begin{equation}
\psi_+ (\vec{k}) = c_{\vec{k}_0 + \vec{k}}~~~,~~~\psi_- (\vec{k}) = c_{-\vec{k}_0 + \vec{k}}.
\end{equation}
Then we expand all terms
in $\mathcal{S}_c + \mathcal{S}_{\phi} + \mathcal{S}_{c \phi}$ in spatial and temporal
gradients. Using the co-ordinate system illustrated in Fig.~\ref{fig:ssfs}, performing appropriate
rescaling of co-ordinates, and dropping terms which can later be easily shown to be irrelevant, we 
obtain the 2+1 dimensional Lagrangian
\begin{eqnarray}
\mathcal{L} &=&  \psi^\dagger_{+ \alpha} \Bigl(\zeta \partial_{\tau} - i  \partial_x - 
 \partial^2_y \Bigr) \psi_{+ \alpha} + \psi^\dagger_{- \alpha} \Bigl(\zeta \partial_{\tau} + i  \partial_x - 
 \partial^2_y \Bigr) \psi_{- \alpha} \nonumber \\ &-& \lambda \, \phi \, \left( 
 \psi^\dagger_{+\alpha} \psi_{+ \alpha} + \psi^\dagger_{- \alpha} \psi_{- \alpha} \right) + \frac{N}{2} (\partial_y \phi)^2 + \frac{N s}{2} \phi^2 . \label{ssl}
\end{eqnarray}
Here $\zeta$, $\lambda$ and $s$ are coupling constants, with $s$ the tuning
parameter across the transition; we will see that all couplings apart from $s$ can be scaled
away or set equal to unity.
We now allow the spin index $\alpha = 1 \ldots N$, as we will be interested in the
structure of the large $N$ expansion. Note that Eq.~(\ref{ssl}) has the same basic structure
as the models considered in Section~\ref{sec:ssdwave}, apart from differences in the 
spatial gradients and the matrix structure. We will see that these
seemingly minor differences will completely change the physical properties and 
the nature of the large $N$ expansion.

\subsection{Symmetries}
\label{sec:sssym}

\begin{figure}[t]
\centerline{\includegraphics[width=1.5in]{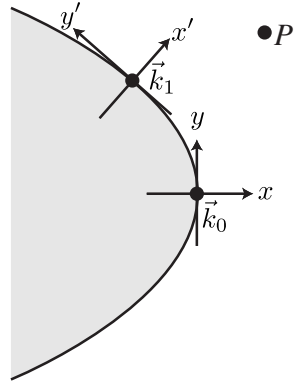}} \caption{
The fermion correlator at the point P can be described either in terms
of the 2+1 dimensional field theory at $\vec{k}_0$, or that at $\vec{k}_1$.} \label{fig:ssshift}
\end{figure}

A first crucial property of $\mathcal{L}$ is that the fermion Green's functions do indeed
have singularities along a line in momentum space, as was required by our discussion
above. This singularity is a consequence of the invariance of $\mathcal{L}$ under the following
transformation
\begin{equation}
\phi(x,y) \to \phi(x,y+\theta x), \quad \psi_s(x,y) \to e^{-i s(\frac{\theta}{2} y + \frac{\theta^2}{4} x)} \psi_s(x,y+\theta x), \label{ssgal}
\end{equation}
where $\theta$ is a constant.
Here we have dropped time and $\alpha$ indices because they play no role, and $s= \pm$.
We can view this transformation as one which performs a `rotation' of spatial co-ordinates,
moving the point $\vec{k}_0$ to neighboring points on the Fermi surface. We are not assuming the
Fermi surface is circular, and so the underlying model is not rotationally invariant. However,
we are considering a limiting case of a rotation, precisely analogous to the manner in which Galilean
transformations emerge as a limiting case of a relativistic transformation ($x$ behaves
like time, and $y$ as space, in this analogy). This `Galilean' symmetry is an emergent symmetry
of $\mathcal{L}$ for arbitrary shapes of the Fermi surface. It is not difficult to now
show from (\ref{ssgal}) that the $\phi$ Green's function $D$, and the fermion
Green's functions $G_s$ obey the exact identities
\begin{eqnarray}
D(q_x, q_y) &=& D(q_y) \label{slideb} \\ 
G_s(q_x, q_y) &=& G(s q_x + q^2_y). \label{slidef}
\end{eqnarray}
So we see that the $\Psi_+$ Green's function depends only on $q_x + q_y^2$. 
The singularities of this function appear when $q_x + q_y^2 = 0$, and this is nothing
but the equation of the Fermi surface passing through the point $\vec{k}_0$
in Figs.~\ref{fig:ssfs} and~\ref{fig:ssshift}. Thus we have established the existence of a line of singularities
in momentum space.

The identities in Eq.~(\ref{ssgal}) also help establish the consistency of our
description in terms of an inifinite number of 2+1 dimensional field theories.
Consider the fermion Green's function at the point P in Fig.~\ref{fig:ssshift}.
This can be computed in terms of the 2+1 dimensional field theory defined at
the point $\vec{k}_0$, or from that at a neighboring point $\vec{k}_1$.
Eq.~(\ref{slidef}) ensures that both methods yield the same result. A little geometry \cite{max1}
shows that $q_x + q_y^2$ is an invariant that measures the distance between
P and the closest point on the Fermi surface: thus it takes the same value in 
the co-ordinates systems at $\vec{k}_0$ and $\vec{k}_1$, with $q_x + q_y^2 = q_x^\prime
+ q_y^{\prime 2} $, establishing the identity of the two computations.

\subsection{Scaling theory}
\label{sec:ssscaling}

Let us now discuss the behavior of $\mathcal{L}$ under renormalization group scaling transformations.
The structure of the spatial gradient terms in the Lagrangian indicates that 
 the rescaling of spatial 
co-ordinates should be defined by
\begin{equation}
x' = x/b^2 \quad, \quad y' =y/b .
\end{equation}
The invariance in Eq.~(\ref{slidef}) implies that these scalings are exact,
and the spatial anisotropy acquires no fluctuation corrections.
Or, in other words
\begin{equation}
\mbox{dim}[y]=-1 \quad, \quad \mbox{dim}[x] = -2 \label{ssdimxy}
\end{equation}
For now, let us keep the rescaling of the temporal co-ordinate general:
\begin{equation}
\mbox{dim}[\tau] = -z . \label{ssdimtau}
\end{equation}
Note that the dynamic critical exponent $z$ is defined relative to the spatial co-ordinate $y$
tangent to the Fermi surface (other investigators sometimes define it relative to the
co-ordinate $x$ normal to the Fermi surface, leading to a difference in a factor of 2).
We define the engineering dimensions of the fields so that co-efficients  of the $y$ derivatives
remain constant. Allowing for anomalous dimensions $\eta_\phi$ and $\eta_\psi$ from loop
effects we have
\begin{equation}
\mbox{dim}[\phi] = (1 + z + \eta_\phi)/2 \quad , \quad \mbox{dim}[\psi] = (1 + z + \eta_\psi)/2 .
\label{ssdimphi}
\end{equation}
Using these transformations, we can examine the scaling dimensions of the couplings in $\mathcal{L}$
at tree level
\begin{equation}
\mbox{dim}[\zeta] =  2-z - \eta_\psi \quad , \quad \mbox{dim}[\lambda] = (3-z - \eta_\phi - 2 \eta_\psi)/2 .
\label{ssdimzeta}
\end{equation}

We will see in Section~\ref{sec:sslargen} 
that low order loop computations suggest that the anomalous dimensions
$\eta_\phi$ and $\eta_\psi$ are small, and that $z \approx 3$. Assuming these estimates
are approximately correct, we see that the coupling $\zeta$ is strongly irrelevant. Thus we can send $\zeta \rightarrow 0$
in all our computations. However, we do not set $\zeta =0$ at the outset, because the temporal derivative
term is needed to define the proper analytic structure of the frequency loop integrals \cite{SSLee}. As we will
discuss later, the limit $\zeta \rightarrow 0$, also dramatically changes the counting 
of powers of $1/N$ in the loop expansion \cite{SSLee}.

Also note that these estimates of the scaling dimensions imply $\mbox{dim}[\lambda] \approx 0$. 
Thus the fermion and order parameter fluctuations remain strongly coupled at all scales.
Conversely, we can also say that the requirement of working in a theory with fixed $\lambda$
implies that $z \approx 3$; this circumvents the appeal to loop computations for taking the $\zeta \rightarrow 0$ limit.
With a near zero scaling dimension for $\lambda$, we cannot
expand perturbatively in powers of $\lambda$. This features was also found in Section~\ref{sec:ssisingnematic},
but there we were able to use the $1/N$ expansion to circumvent this problem. 

Moving beyond tree level considerations, we note that another Ward identity obeyed by the theory
$\mathcal{L}$ allows us to fix the scaling dimension of $\phi$ exactly. This Ward identity is linked to the
fact that $\phi$ appears in the Yukawa coupling like the $x$ component of a gauge field coupled to the
fermions \cite{max1}. The usual arguments associated with gauge invariance then imply 
that $\mbox{dim}[\phi] = 2$ (the same as the
scaling dimension of $\partial_x$), and 
that we can work in theory in which the ``gauge coupling'' $\lambda$ set equal to unity at all scales.
Note that with this scaling dimension, we have the exact relation
\begin{equation}
\eta_\phi = 3-z .
\end{equation}
Note also that Eq.~(\ref{ssdimphi}) now implies that $\mbox{dim}[\lambda] = \eta_\psi$ at tree level, 
which is the same
as the tree level transformation of the spatial derivative terms. The latter terms have been set equal
to unity by rescaling the fermion field, and so it is also consistent to set $\lambda =1$ from now on.

We reach the remarkable conclusion that at the critical point $s=s_c$, $\mathcal{L}$ is independent of 
all coupling constants. The only parameter left is $N$, and we have no choice but to expand
correlators in powers of $1/N$. The characterization of the critical behavior only requires computations
of the exponents $z$ and $\eta_\psi$, and associated scaling functions.

We can combine all the above results into scaling forms for the $\phi$ and $\Psi$ Green's
functions at the quantum critical point at $T=0$. These are, respectively
\begin{eqnarray}
D^{-1} (q_x, q_y, \omega) &=& q_y^{z-1} \mathcal{F}_D \left( \frac{\omega}{q_y^{z}} \right) 
\label{ssdi2} \\
G^{-1} (q_x, q_y, \omega)  &=& (q_x + q_y^2)^{1 - \eta_\psi/2} \mathcal{F}_G \left(
\frac{\omega}{(q_x + q_y^2)^{z/2}} \right), \label{ssgi1}
\end{eqnarray}
where $\mathcal{F}_D$ and $\mathcal{F}_G$ are scaling functions.

\subsection{Large $N$ expansion}
\label{sec:sslargen}

We have come as far as possible by symmetry and scaling analyses alone on $\mathcal{L}$.
Further results require specific computations of loop corrections, and these can only be carried
out within the context of the $1/N$ expansion. At $N=\infty$, the $\phi$ propagator at criticality is
\begin{equation}
\frac{D^{-1}}{N} = q_y^2 + \frac{|\omega|}{4 \pi |q_y|}
\label{ssdi1}
\end{equation}
for imaginary frequencies $\omega$. This is clearly compatible with Eq.~(\ref{ssdi2}) 
with $z=3$. The leading correction to the fermion propagator comes from the self energy
associated with one $\phi$ exchange, and this leads to
\begin{equation}
G_+^{-1} = -i \zeta \omega + q_x + q_y^2 - i \mbox{sgn} (\omega) \frac{2}{\sqrt{3} (4 \pi)^{2/3} N }
|\omega|^{2/3}, \label{ssgi2}
\end{equation}
which is also compatible with Eq.~(\ref{ssgi1}) with $z=3$ and $\eta_\psi = 0$.
Notice also that as $\omega \rightarrow 0$, the $\zeta \omega$ term in 
Eq.~(\ref{ssgi2}) is smaller than the $|\omega|^{2/3}$ term arising from the self energy
at order $1/N$; this relationship is equivalent to our earlier claim that $\zeta$ is irrelevant
at long scales, and so we should take the limit $\zeta \rightarrow 0$ to obtain our
leading critical scaling functions.

The structure of Eq.~(\ref{ssgi2}) also illustrates a key difficulty associated with the
$\zeta \rightarrow 0 $ limit. At $\zeta = 0$, the leading $\omega$ dependence of 
$G^{-1}$ is $\sim |\omega|^{2/3}/N$. Feynman graphs which are sensitive to this $\omega$
dependence will therefore acquire additional factors of $N$, leading to a breakdown
of the conventional counting of powers of $1/N$ in the higher loop graphs.

This breakdown of the $1/N$ expansion 
was investigated by Lee \cite{SSLee} for a `single patch' theory with fermions only at $\vec{k}_0$
(and not at $- \vec{k}_0$). We see from Eq.~(\ref{ssgi2}) that the $1/N$ term in $G^{-1}$ 
becomes important when $q_x = q_y = 0$ {\em i.e.\/} the fermion is precisely on the Fermi surface.
Thus the power of $N$ is maximized when fermions in all internal lines are on the Fermi surface.
Such a Fermi surface restriction is satisfied only in a subspace of reduced dimension in 
the momentum space integral of any Feynman graph. Lee presented an algorithm for computing
the dimensionality of this restricted subspace: he demonstrated that the power of $1/N$ was
determined by the genus of the surface obtained after drawing the graph in a double-line representation.
So determining the leading $1/N$ terms in Eqs.~(\ref{ssdi1}) and (\ref{ssgi2}) requires summation
of the infinite set of planar graphs. This problem remains unsolved, but the unexpected
appearance of planar graphs does suggest an important role for the AdS/CFT correspondence.

The structure of the loop expansion for the `two patch' theory with fermions at $\pm \vec{k}_0$,
as written in Eq.~(\ref{ssl}), was studied in Ref.~\cite{max1}. It was found that $z=3$ was preserved
upto three loops, but a small non-zero value for $\eta_\psi$ did appear at three loop order.
Also, the genus counting of powers of $1/N$ was found to break down, with larger
powers of $N$ appearing in some three loop graphs.

\subsection{AdS/CFT correspondence}
\label{sec:ssadsmetal}

There has been a great deal of recent work \cite{sslee3,schalm,lmv,ssdenef,ssshankar,sssinha,sskraus,sssandip,ssgary,ssryu,ssgubser,ssklebanov,sshartman,sstong,ssjohnson,sspolchinski,sskarch} investigating 
the structure of Fermi 
surfaces using the AdS/CFT correspondence. The results obtained so far do
have features that resemble our results above for the Ising-nematic transition
in two dimensional metals. However, the precise connection remains obscure,
and is an important topic for future research. In particular, a microscopic understanding
of the field content of the CFT dual of the AdS theory is lacking, although there
has been interesting progress very recently \cite{ssklebanov,sspolchinski}.

One of the main results of the analysis of Ref.~\cite{lmv} is the general
structure of the fermion Green's function obtained in a theory dual to a Reissner-Nordstrom
black hole in AdS$_4$. This had the form
\begin{equation}
G^{-1} ( \vec{k}, \omega)  = - i \omega + v_F ( |\vec{k}| - k_F ) - c_1 \omega^\theta ,
\label{ssadsferm}
\end{equation}
where the momentum $\vec{k}$ is now measured from the origin of momentum space (and not
from a Fermi surface), and the complex number $c_1$ and exponent $\theta$ are computable
functions of the ultraviolet scaling dimension of the fermion field. The AdS theory only considers a circular
Fermi surface, and for this geometry (after appropriate rescaling)
\begin{equation}
v_F ( |\vec{k}| - k_F) = v_F ( |\vec{q} + \vec{k}_0| - k_F ) \approx q_x + q_y^2 ;
\end{equation} 
now Eq.~(\ref{ssadsferm})
is seen to be strikingly similar to Eq.~(\ref{ssgi2}).
Ref.~\cite{lmv} also argued that Eq.~(\ref{ssadsferm}) was a generic property of the 
near horizon geometry of the Reissner-Nordstrom black hole: the geometry 
changes from AdS$_4$ near the boundary to AdS$_2 \times R^2$ near the black hole horizon.

It is interesting to compare the structure of the critical theory in the AdS/CFT framework
to that found in the subsections above for the Ising-nematic transition in a metal.
The latter was described by an infinite set of 2+1 dimensional field theories labeled by pairs of momenta
on a one-dimensional Fermi surface {\em i.e.\/} a $S^1 /Z_2$ set of 2+1 dimensional field theories.
In the low-energy limit, the AdS/CFT approach yields \cite{lmv} a AdS$_2 \times {R}^2$ geometry: 
this can be interpreted
as an infinite set of chiral 1+1 dimensional theories labeled by a ${R}^2$ 
set of two-dimensional momenta
$\vec{k}$. It is notable, and perhaps significant, that both approaches have an emergent 
dimension not found in the underlying degrees of freedom.
The Ising nematic theory 
began with a 2+1 dimensional Hamiltonian $\mathcal{S}_c + \mathcal{S}_\phi + \mathcal{S}_{c \phi}$ in
Eqs.~(\ref{g15},\ref{sc},\ref{sci}), and ended up with a ${S}^1 /{Z}_2$ set of 
2+1 dimensional field theories. In AdS/CFT, there is the emergent radial direction representing energy scale.
These emergent dimensions imply redundant descriptions, and require associated consistency conditions:
we explored such consistency  conditions in Section~\ref{sec:sssym}, while in AdS/CFT the
consistency 
conditions are Einstein's equations representing the renormalization group flow under changes of energy scale.
It would be interesting to see if fluctuations about the classical gravity theory
yield corrections to the AdS$_2 \times {R}^2$ geometry which clarify the connection to our 
Ising-nematic theory.

\subsubsection*{Acknowledgements}

Many of the ideas reviewed in Section~\ref{sec:ssmetal} are due to Max Metlistki \cite{max1}; I also thank him for
valuable comments on the manuscript.
I thank the participants of the school for their interest, and for stimulating discussions.
This research was supported by the National Science Foundation under grant DMR-0757145, by the FQXi
foundation, and by a MURI grant from AFOSR.

\end{document}